\documentclass[twocolumn,showpacs,preprintnumbers,amsmath,amssymb,floatfix]{revtex4}
\usepackage{amssymb,amsmath,eepic,epsfig,psfrag}
\usepackage[colorlinks=true,citecolor=blue,linkcolor=blue]{hyperref}
\usepackage{dcolumn}
\usepackage{graphics}
\usepackage{graphicx}
\usepackage{subfigure}
\usepackage{epsfig}
\usepackage{setspace}
\usepackage{color}

\def \be {\begin{equation}}
\def \ee {\end{equation}}
\def \ba {\begin{eqnarray}}
\def \ea {\end{eqnarray}}
\def \bm {\begin{displaymath}}
\def \em {\end{displaymath}}
\def \br {{\bf r}}

\newcommand{\Rnum}[1]{\uppercase\expandafter{\romannumeral #1\relax}}
\begin{document}
\title{ Cassie - Wenzel transition of a binary liquid mixture on a nano-sculptured surface}
\author{Swarn Lata Singh{$^{1,2}$}, Lothar Schimmele{$^{1,2}$}, and S. Dietrich{$^{1,2}$}}
 \affiliation{$^{1}$Max-Planck-Institut f\"{u}r
 Intelligente Systeme, D-70569 Stuttgart, Heisenbergstr. 3, Germany, \\$^{2}$ Institut f\"{u}r
 Theoretische  Physik, Universit\"{a}t Stuttgart, Pfaffenwaldring 57,
 D-70569 Stuttgart, Germany
 }
\date{\today}
\begin{abstract}
The Cassie - Wenzel transition of a symmetric binary liquid mixture in contact with a
nano-corrugated wall is studied. The corrugation consists of a periodic array of  nano-pits
with square cross sections. The substrate potential is the sum over Lennard-Jones interactions,
describing the pairwise interaction between the wall particles $C$ and the fluid particles.
The liquid is composed of two species of particles, $A$ and $B$, which have the same size and
equal $A - A$ and $B - B$ interactions. The liquid particles interact between each other also
via $A - B$ Lennard-Jones potentials. We have employed classical density functional theory
to determine the equilibrium structure of binary liquid mixtures in contact with the
nano-corrugated surface. Liquid intrusion into the pits is studied as a function of various
system parameters such as the composition of the liquid, the strengths of various inter-particle
interactions, as well as the geometric parameters of the pits. The binary
liquid mixture is taken to be at its mixed-liquid-vapor coexistence. For various sets of
parameters the results obtained for the Cassie - Wenzel transition, as well as for the metastability
of the two corresponding thermodynamic states, are compared with macroscopic predictions
in order to check the range of validity of the macroscopic theories for systems exposed
to nanoscopic confinements. Distinct from the macroscopic theory, it is found that
the Cassie - Wenzel transition cannot be predicted based on the knowledge of a single parameter,
such as the contact angle within the macroscopic theory.
\end{abstract}
\maketitle

\pagebreak

\section{\bf Introduction}

Wetting of solid surfaces by liquids are ubiquitous in nature \cite{neinhuis, waterstrider, Genzer}
as well as in various technological applications \cite{Wang, Blossey, Delamarche,Bhushan, 
David,Heuberger}. It is well established that topography,
both geometrical and chemical, plays an important role in controlling
the wettability of a surface \cite{Alberto, Jayant1com}.
With the advent of advanced fabrication techniques, it is possible
to tailor the topography of a surface down to the nano-scale \cite{mp9, mp10}.
One observes that changing the topography, even only on the nano-scale, results
in large changes of macroscopic observables, such as 
the contact angle of a sessile drop. This has encouraged the fabrication
of patterned surfaces in order to control wetting. Typical configurations,
which liquids exhibit on textured surfaces, are either the Cassie-Baxter (CB)
\cite{CB} or the Wenzel (W) state \cite{Wenzel}. In the CB state liquid
remains suspended above the substrate surface, whereas in the Wenzel
state the liquid intrudes the surface cavities, {\it i.e.}, pits.

The requirement of surface wettability, however,
varies from one application to the other. Several manufacturing,
biological, and agrochemical applications require the liquid to
spread rapidly on the surface \cite{Ivanova, Iv4, Iv6, Zhang17},
whereas poor wettability is required in designing liquid-repellent
surfaces \cite{Butt1, Jayant6, Jayant8}.
Modifying the surface properties is the most exploited method to 
achieve the desired wettability for a given system. 
In addition to changing the surface texture, adding a surfactant 
to the liquid in order to modify its wetting behavior 
provides additional flexibility. Surfactants are widely
used to facilitate wetting of surfaces in many industrial
applications such as agrochemical, textile, chemical, and pharmaceutical
industries \cite{Zhang, Jayant30, Jayant31, Jayant32}.
In certain circumstances, it is not easy to change the surface
properties, so  blending the liquid turns out to be 
a more feasible option to control the wettability. 
Enhancing pesticide utilization on plant leaves is such an example, 
in which surfactants are added to the pesticides in order to 
enhance their spreading on leaves
\cite{Zhang17, Ivanova, Zhang11, Zhang13, Zhang14}.

Although wetting of solid surfaces by a liquid mixture
is  encountered frequently in many industrial applications 
and biological systems, most of the actual studies concerning 
wetting focus on one$-$component liquids. There have been very 
few experimental studies regarding the wetting of a solid surface
by liquid mixtures \cite{Ivanova, Zhang, Jayant17, Jayant18, Plech} and also
only very few theoretical investigations 
\cite{pd1, pd2, Jayant, Neil, Jayant28, Tanaka, PRA91} thereof.
A microscopic understanding of the {\mbox{Cassie $-$}}
Wenzel transition for liquid mixtures 
is crucial for many application purposes.
Here, we present a study of the Cassie $-$
Wenzel transition of a binary liquid mixture at a corrugated surface.
The binary liquid mixture is composed of two types of particles 
labeled as A and B particles.
For simplicity the studies here are limited to symmetric liquids, 
{\it i.e.}, the radii of the A and B  particles are the same,
as well as the strengths of the A $-$ A and
B $-$ B interactions,
whereas the A $-$ B interaction is varied. The wall is composed of 
C particles and modeled as  a periodically repeated array of 
nano-pits with square cross sections of width $w$. The depth 
of the pit is denoted as $D$ and its side walls are vertical. 
Without loss of generality, we consider the case that the 
attractive B $-$ C interaction is stronger
than the A $-$ C interaction. Here we use classical
density functional theory (DFT) as a tool to gain a  microscopic
understanding of the wetting transition. In order to reduce the number
of independent thermodynamic variables, the following discussions are limited to 
thermodynamic conditions for which the liquid and vapor 
phases coexist in the bulk.

Young's contact angle $\theta_{Y}$ characterizes macroscopically 
the wetting behavior of a liquid on a planar and 
homogeneous solid surface:

\be \label{eq:1}
\cos\theta_{Y} = \frac{\sigma_{sv} - \sigma_{sl}}{\sigma_{lv}},
\ee
where $\sigma_{sv}$, $\sigma_{sl}$, and $\sigma_{lv}$ are
the interfacial  tensions of the solid-vapor, solid-liquid, and
liquid-vapor interface, respectively.
According to macroscopic theory, a liquid, which is at bulk 
liquid-vapor coexistence and is brought in contact with a 
structured wall, as described above consisting of square 
pits with vertical sides, remains in the Cassie state if 
$\theta_{Y} > 90^{\circ}$, whereas for
$\theta_{Y} < 90^{\circ}$ the liquid intrudes the pits.
Within this description, the Cassie-Wenzel transition 
occurs at $\theta_{Y} = 90^{\circ}$,
irrespective of which of the various system parameters has been 
tuned in order to obtain a contact angle of $90^{\circ}$.

Apart from the Cassie to Wenzel transition (or intrusion 
transition), information about the metastability of the Cassie
and Wenzel states is of high interest as well.
The metastability of a liquid on a given surface could be
exploited for drug delivery and for designing 
superhydrophobic or omniphobic surfaces \cite{Alberto, Alberto2}.
If, for given thermodynamic conditions, the Cassie state
is not stable on the surface, but  remains metastable for
a desired duration, de facto superhydrophobicity is achieved.
This would diminish the challenges to design
superhydrophobic surfaces.
The macroscopic equation for the coexistence of the Cassie and
the  Wenzel state on a textured surface yields, at bulk liquid-vapor 
coexistence, a  critical contact angle $\theta^{c}_{Y}$ 
at which both states have the same grand canonical potential 
such that the Wenzel state is stable for 
$\theta_{Y} < \theta^{c}_{Y}$ and the Cassie 
state is stable for $\theta_{Y} > \theta^{c}_{Y}$. This 
critical contact angle depends only on the
geometrical parameters of the pits, and not on
details of the liquid properties. Within a  macroscopic
description, the relative stability of the Cassie and the  Wenzel
state is controlled entirely by a single macroscopic parameter,
{\it i.e.}, the contact angle $\theta_{Y}$. Although the same $\theta_{Y}$
can be achieved with a multitude of combinations of system 
parameters, on a macroscopic level all these details are 
irrelevant concerning the issue of the relative 
stability of the Cassie and the Wenzel state. 
For a surface endowed with pits of square cross section of
width $w$ and depth $D$ as well as with vertical side walls,  
at liquid-vapor coexistence $\theta^{c}_{Y}$ 
can be derived from macroscopic theory as outlined 
in Sec. \Rnum{4} (see Eq. (\ref{eq:17}); a similar 
expression for grooves was given in Ref. \cite{David}) :

\be \label{eq:2}
\cos\theta^{c}_{Y} = - \frac{w}{4D + w}. 
\ee

For deep pits, {\it i.e.}, $D/w \to \infty$, one has 
$\theta^{c}_{Y} = 90^{\circ}$. However, for shallow pits
$\theta^{c}_{Y}$ may attain 
values considerably larger than $90^{\circ}$. 
For  contact angles $\theta_{Y} < \theta^{c}_{Y}$,
the Wenzel state is stable. However, if $\theta^{c}_{Y} > 90^{\circ}$,
macroscopic theory tells  that within the range  
$90^{\circ} \leq \theta_{Y} < \theta^{c}_{Y}$ the Cassie state 
is still metastable, because the liquid can intrude the 
pits only if $\theta_{Y} < 90^{\circ}$. For $\theta_{Y} < 90^{\circ}$, 
the Cassie state is  unstable.
For $\theta_{Y} > \theta^{c}_{Y}$ the Cassie state is 
stable and the Wenzel state is metastable up to 
a sufficiently large contact angle (depending on the shape of
the pits) above which it is unstable. 
For a pit with rectangular corners and at bulk 
fluid-vapor coexistence, the critical angle, above which 
the Wenzel state is {\it u}nstable, is
given by $cos\theta^{c}_{u} = -\frac{1}{\sqrt{3}}$
({\it i.e.}, $\theta_{Y}\approx 125^{\circ}$). The value of this
critical angle follows from simple geometric considerations,
which amount to construct a planar fluid$-$vapor 
interface meeting the three walls forming the corner 
at the bottom at the angle $\theta_{Y}$ 
\hspace{0.01cm} (independent of whether the fluid is multicomponent or not).

The aim of the present study is to test to which extent 
the above mentioned macroscopic predictions concerning the 
Cassie $-$ Wenzel transition are valid for liquids on a 
nano $-$ corrugated surface,
with a focus on effects specific to multicomponent fluids.
In general, for fluids in nano-confinement
a macroscopic description becomes unreliable. Therefore it is 
expected that the phenomena discussed above are no longer 
controlled by a single macroscopic parameter, such as the contact angle.
Instead, one expects that more details like the fluid composition,
the fluid $-$ fluid, and the fluid $-$ wall interactions eventually become
relevant. Therefore, we have studied the Cassie $-$ Wenzel
transition of a binary liquid mixture on a nano-textured wall by 
using density functional theory (DFT). We have analyzed 
extensively the intrusion as a function of the properties
of the liquid
(e.g, the composition of the liquid and the relative strengths
of the A $-$ A and A $-$ B interactions), in order
to assess the effect of these properties on the Cassie $-$ Wenzel
transition and on the relative stabilities of the two states.
The other control parameters considered in this study are
the strengths of the fluid-solid  
interactions (A $-$ C and B $-$ C) 
of the two fluid species and the dimensions
of the pit. We have varied the dimensions of the pit in order to study the 
effect of strong confinement on the wetting behavior. 
For each set of parameters, we have calculated the number density
profiles for both types of fluid particles for the stable and, if
appropriate, also the metastable configurations. Based on them 
we have monitored the intrusion at
bulk liquid-vapor coexistence.
In order to be able to compare the microscopic results with the macroscopic
predictions, we have calculated the interfacial tensions for 
each model studied via DFT as well as the corresponding
macroscopic Young contact angle based on Eq.  (\ref{eq:1}).
The critical contact angle $\theta_{Y}^{c}$, below which the intrusion is 
observed within DFT, is compared with the macroscopic predictions.
In addition, in order to determine the relative stability of the Cassie 
and the Wenzel states we have calculated the grand canonical
potential $\Omega$ for each number density configuration generated by DFT. 
The contact angle corresponding to the coexistence of the Cassie and 
Wenzel states is compared with $\theta^{c}_{Y}$ given by Eq.  (\ref{eq:2}).

The paper is organized as follows. In Sec. \Rnum{2} we provide
a detailed description of the system studied here. It is characterized completely by
the texture of the surface as well as by the fluid-fluid and
fluid-surface interactions. In Sec. \Rnum{3} we give a brief
introduction to the  technique used for the present
investigation. In Sec. \Rnum{4} we discuss our results
and conclude in Sec. \Rnum{5}.
\section{Fluid-fluid interaction, wall topography, and substrate potential}

\begin{figure}[h!]     
\centering
\hspace*{-0.0cm}\includegraphics[scale=0.40]{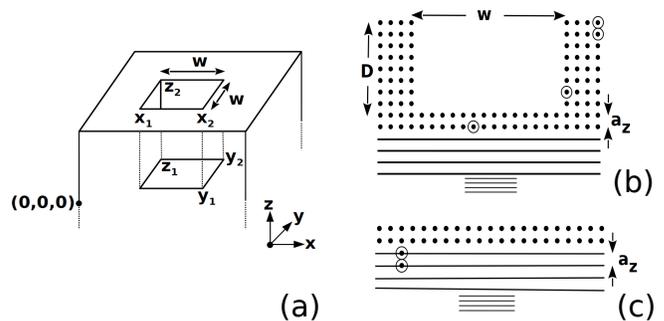}%
{\vspace*{-0.0cm}
\caption{\label {fig1} 
$(a)$ Schematic representation of the structured wall studied 
here. The structure is modeled by square pits $w \times w \times D$
of width $w$ and depth $D$. The pits are carved out from a planar
wall. In $(a)$ and $(b)$ a single pit is shown. If the pit and a 
neighboring part of the wall are considered as a unit cell, which is repeated 
periodically in $x$ and $y$ direction, one obtains a two-dimensional lattice of
the pits. The solid substrate is modeled as a simple cubic lattice 
occupied densely by particles of type C. The lattice spacing is 
$a_{z}$ which equals twice the radius $R_{C}$ of the
C particles. The circles indicate the size of the C 
particles, which are densely packed.
In $(b)$, the cross section of the pit is shown. For technical convenience
the horizontal lattice planes below the two forming the bottom of 
the pit are taken to be the lateral average in $x$ and $y$
direction of the discrete lattice structure, maintaining the 
discrete vertical structure. These homogeneous planes are 
indicated by horizontal lines; they form a half space.
The liquid in the pit is sufficiently far from these 
homogeneous lattice planes so that their missing lateral 
corrugation is quantitatively negligible for the substrate
potential. On the other hand, keeping the full lattice structure 
at the top, at the sidewalls, and at the bottom of the pit gives 
rise to the same local contact angle. In $(c)$ the reference 
configuration of the planar substrate is shown, for which 
$\sigma_{sv}$ and $\sigma_{sl}$ (see Eq. (\ref{eq:1}))
are determined. 
The lengths $w$ and $D$ are measured between the respective
loci of the nuclei of the C praticles. These definitions
imply that they encompass the corresponding depletion zone 
for the density profiles of the fluid particles. 
}
}
\end{figure}

We have studied binary liquid mixtures in contact with a 
nano-structured wall at the bottom and the bulk
of the liquid at the top. The wall is endowed with a periodic 
array of square nano-pits, with edge length $w$ and 
depth $D$ (see  Fig. $1$). The liquid is composed
of A and B particles. These particles are considered to interact via Lennard-Jones (LJ)
pair potentials. In the spirit of DFT as used here and discussed in Sec. \Rnum{3}, each 
pair potential is replaced by the sum of a hard sphere interaction

\begin{equation} \label{eq:3}
U^{hs}_{ij}(r) =   \left\{
\begin{array}{r l}
\infty,  &  r\leq\sigma_{ij} = R_{i} + R_{j} , \\ 
 0,      &  r > \sigma_{ij}
\end{array}
\right.
\end{equation}
and of a soft attractive part \cite{Weeks}
\begin{eqnarray}\label{eq:4}
U^{att}_{ij}(r)=-\epsilon_{ij} \Theta({2}^{1/6}\sigma_{ij}-r)
+\Phi^{LJ}_{ij}(r)\Theta(r-{2}^{1/6}\sigma_{ij}) 
\end{eqnarray}
with the Heaviside function $\Theta$ and
\begin{equation}\label{eq:5}
\Phi^{LJ}_{ij}=4\epsilon_{ij}\left[{\left(\frac{\sigma_{ij}}{r}\right)}^{12}-{\left(\frac{\sigma_{ij}}{r}\right)}^{6}\right]. 
\end{equation}

Here, $i$ and $j$ represent the two species A and B of 
the binary liquid mixture, $R_{i}$ is the radius of species $i$ of the fluid particles,
$-\epsilon_{ij}$ is the potential depth for the $ij$-pair 
potential at $r=2^{1/6}\sigma_{ij}$, $r$ is the center-to-center interparticle separation,
and $\sigma_{ij}$ is the distance of contact between the centers of two interacting 
liquid particles. For simplicity, the binary liquid mixture is considered to be symmetric
with $R_{A} = R_{B}$ and $\epsilon_{AA} = \epsilon_{BB}$,
where $\epsilon_{ij}$ is the parameter of the interaction strength
of the $ij$-pair potential.
The fluid-fluid interaction is rendered effectively
short-ranged by introducing a cut-off at $r=R_{c}$.
In the following we adopt $R_{c} = 5R_{A} = 5R_{B}$ which is
implemented by a cut-off function. 
The interaction parameters $\epsilon_{ij}$ for the cut-off
potential are rescaled such that the integrated interaction is equal 
to the original one resulting from the potential in Eq. (\ref{eq:5}) with $R_{c} \to \infty$.
For further details see Ref. \cite{mypre}.

The model for the corrugated wall is introduced and described in Fig. $1$. 
According to Fig. $1$, the  very bottom of the wall consists of laterally 
homogenous layers (see the horizontal lines in Fig. $1$) which 
form a half-space (see below). These layers exhibit an interlayer 
spacing $a_{z} = 2R_{c}$. 
The layers are of macroscopic extent in  the lateral $(x, y)$ 
directions. 
This half-space attracts the fluid particles via van der Waals 
forces A $-$ C and B $-$ C:

\begin{equation}\label{eq:6}
U^{\rm half-space}_{i}(r)=-4\epsilon_{i}\sum_{l}{\left[\frac{{{\sigma_{i}}^{2}}}{{|{\bf r}-{{\bf r}_{l}}|}^{2}}\right]}^{3} ,
\end{equation}
where $i=A,B$ is the fluid particle at the position $\bf r$, 
$l$ represents a half-space particle at position ${{\bf r}_{l}}$, 
$\sigma_{i}$ is a (microscopic) length which can be chosen 
to provide the units of the number densities. 
In combination with the energy parameter $\epsilon_{i}$, it describes 
the strength of the interaction between a fluid particle of type
$i$ and a wall particle. Since the layers forming the half-space
are homogeneous in the lateral directions, 
the summation in Eq. (\ref{eq:6}), in the $x$ and $y$ directions 
reduces to an integration for each layer, 
and the contributions from the layers 
are summed up. For further details, we refer to Ref. \cite{mypre}.

The solid particles C occupying the remaining fully discrete, simple cubic lattice sites
on top of the half $-$ space interact with the liquid particles via a Lennard-Jones
potential: 
\begin{equation}\label{eq:7}
U^{\rm structure}_{i}(r) = -4\epsilon_{i}\sum_{l}{\left[{\left(\frac{\sigma_{i}}{|{\bf r}-{{\bf r}_{l}}|}\right)}^{12} 
- {\left(\frac{\sigma_{i}}{|{\bf r}-{{\bf r}_{l}}|}\right)}^{6}\right]} .
\end{equation}
The parameters $\epsilon_{i}$ of the interaction strengths
for the solid-fluid pair interaction are the same as 
the parameters of the interaction strengths for the half-space--liquid 
interaction. The same holds for $\sigma_{i}$,
where $\sigma_{i} = R_{i} + R_{C}$, $R_{i}=R_{A}=R_{B}$ is the radius of
the liquid particles, and $R_{C}$ is the radius
of the solid particles. 
In Eq. (\ref{eq:7}) the sum over $l$ amounts to the sum over all remaining discrete 
lattice sites occupied by C particles. 

The total fluid-solid (fs) interaction, is the sum of both contributions:

\begin{equation}\label{eq:8}
U^{fs}_{i}(r) = U^{\rm half-space}_{i}(r) + U^{\rm structure}_{i}(r),
\end{equation}
where $i=A,B$ represents the two species of the binary liquid mixture.
\section{Density Functional Theory}
The grand canonical potential $\Omega$ of a classical system
of an $N$-component mixture follows from a variational functional 

\begin{equation}\label{eq:9}
\Omega[\{\rho_{i}\}]=F[\{\rho_{i}\}]+\sum_{i=1}^{N}\int d^{3}r \rho_{i}({\bf r})(V_{i, ext}({\bf r})-\mu_{i}) 
\end{equation}
of the one-particle number densities $\rho_{i}({\bf r}), i = 1,...., N$.
$F$ is the free energy functional, $V_{i, ext}({\bf r})$
is the external potential, and $\mu_{i}$ is the chemical potential of 
species $i=A,B$, respectively. The equilibrium number densities $\rho_{i, 0}
({\bf r})$ minimize $\Omega$:

\begin{equation}\label{eq:10}
\left. \frac{\delta \Omega[\rho_{i}]}{\delta \rho_{i}({\bf r})}\right\vert_{\rho_{i}(\br)=\rho_{i, 0}(\br)}=0 .
\end{equation}
$\Omega[\{\rho_{i, 0}\}]$ is the equilibrium grand canonical potential of the system \cite{Evans2, gurug}.
The free energy functional consist of two parts:

\begin{equation}\label{eq:11}
F[\{\rho_{i}\}]=F_{id}[\{\rho_{i}\}]+F_{ex}[\{\rho_{i}\}],
\end{equation}

where $F_{id}$ is the ideal gas part

\begin{equation}\label{eq:12}
F_{id}[{\rho_{i}}]=k_{B} T \sum_{i=1}^{N} \int d^{3}r \rho_{i}(\br)\left[\ln\left(\rho_{i}(\br)\Lambda_{i}\right)-1\right],
\end{equation}
$\Lambda_{i}={(\frac{h^{2}}{2\pi m_{i} k_{B}T})}^{3/2}$ is
the cube of the thermal wavelength associated with a particle 
of species $i$ and mass $m_{i}$, $h$ is Planck's constant, and 
$k_{B}$ is the Boltzmann constant.

The excess part $F_{ex}$ arises due to the interparticle 
interactions. We approximate the excess part as the 
sum of two distinct contributions: one
arising due to the hard core repulsion $(F_{hs})$, and the 
other due the attractive part of the interaction
$(F_{att})$:

\begin{equation}\label{eq:13}
F_{ex}=F_{hs}+F_{att}.
\end{equation}
$F_{hs}$ is treated within the framework of fundamental measure 
theory (FMT), as described in the next section. $F_{att}$ is 
approximated within a simple random phase approximation.
\subsection{Fundamental Measure Theory}
The fundamental-measure excess free-energy functional for a mixture 
of hard spheres, as proposed by Rosenfeld, is given by
\cite{Roth3, Tarazona, Yasha1}:

\begin{equation}\label{eq:14}
\beta F_{hs}[\rho_{i}]=\int d^{3} r \hspace{0.1cm} {\phi(\{n_{\alpha}({\bf r})\})}
\end{equation}
where the excess free energy density $\phi$ is a function
of the weighted densities $n_{\alpha}({\bf r})$, defined as

\begin{equation}\label{eq:15}
n_{\alpha}({\bf r})=\sum_{i=1}^{N}\int d^{3} r{^{'}}\rho_{i}({\bf r}^{'}) \omega_{i, \alpha}({\bf r}-{\bf r}^{'}), 
\end{equation}
where the weight functions $\omega_{i, \alpha}$ characterize the geometry of the 
spherical particles of species $i$. These weight functions are  
given by \cite{Roth3, Roth4} : 

\begin{eqnarray*}
&&\omega_{i, 3}({\bf r})=\Theta(R_{i}-r), \hspace{0.5cm} \omega_{i, 2}({\bf r})=\delta(R_{i}-r),\\ \nonumber 
&&{{\boldsymbol \omega}}_{i, 2}({\bf r})=\frac{\bf r}{r}\delta(R_{i}-r), \hspace{0.5cm} \nonumber 
\omega_{i, 1}({\bf r}) = \frac{\omega_{i, 2}({\bf r})}{4\pi R_{i}}, \\ \nonumber
&&{\boldsymbol{\omega}}_{i, 1}({\bf r})=\frac{{{\boldsymbol{\omega}}_{i, 2}} ({\bf r})}{4\pi R_{i}}, 
\hspace{0.2cm} {\mbox{and}} \hspace{0.2cm} \omega_{i, 0}({\bf r})=\frac{\omega_{i, 2}({\bf r})}{4\pi R_{i}^{2}}, \\ \nonumber
\end{eqnarray*}

where $R_{i}$ is the radius of the spherical particles of 
species $i$, $\Theta$ is the Heaviside step function, 
and $\delta$ is the Dirac delta function.

In the present study, a modification of the original Rosenfeld 
functional, as proposed by Rosenfeld {\it et al.} \cite{Yasha2}
and known as the modified Rosenfeld functional (MRF), 
has been used. Implementing this functional avoids spurious singular behaviors
in situations with sharply peaked density distributions which 
might emerge if the original Rosenfeld functional \cite{Roth3, Yasha1} 
or refined versions thereof \cite{HGoos} are used. 
On the other hand, the high
accuracy of these functionals is largely preserved for the
systems studied here.
The free energy density $\phi$ within the MRF framework is
\begin{eqnarray}\label{eq:16}
\phi=&&-n_{0}\ln(1-n_{3})+\frac{n_{1}n_{2}-{{{\bf n}_{1}} \cdot {{\bf n}_{2}}}}{1-n_{3}} \\ \nonumber 
&&+\frac{({n_{2})}^{3}}{24\pi{(1-n_{3})}^{2}}{[1-{{{\bf \xi}}}^{2}]}^{q},
\end{eqnarray}
where $q\ge 2$ and ${\boldsymbol \xi}=\frac{{{\bf n}_{2}}}{n_{2}}$ (note that
${{\bf n}_{2}} \cdot {{\bf n}_{2}} \neq {{(n_2)}^2}$).
We have chosen $q=3$ which reproduces the original Rosenfeld functional up 
to the order ${\boldsymbol \xi}^{2}$. In order to approximate the contribution to 
the free energy due to the attractive part of the interaction, we have used
the following truncation of the corresponding functional perturbation expansion:
\begin{equation}
F_{att}=\frac{1}{2}\sum_{i,j = 1}^{N}\int d^{3} r \int d^{3}r^{'}\rho_{i}({\bf r})\rho_{j}({\bf r^{'}})U^{att}_{ij}({\bf r}-{\bf r^{'}}), \nonumber
\end{equation}
with $U^{att}_{ij}$ defined via Eqs. (\ref{eq:4}) and (\ref{eq:5}). Here, the 
minimization of  $\Omega[\{\rho_{i}\}]$
has to be carried out numerically. The number density 
is discretized on a regular simple cubic grid, and a Piccard 
iteration scheme is used in order to minimize $\Omega$ and to 
determine the equilibrium number densities. 
The weighted densities are calculated 
in Fourier space using the convolution theorem. Further
details about these techniques 
can be found in Refs. \cite{mypre, Roth4}. The distributions $\delta$ and $\Theta$
have been smeared out in order to achieve stable convergence for the convolution
(for details see Ref. \cite{mypre}). 

\section{Computational details and Results}
In this section we present the results of DFT calculations
concerning the intrusion of a binary liquid mixture 
at liquid-vapor coexistence in the bulk into the pits 
of a corrugated wall. These calculations have been 
carried out in a computational box, the linear dimensions
of which in $x$, $y$, and $z$ direction have been chosen 
as $26R_{A} \times 26R_{A} \times 40R_{A}$
(with the radii $R_{A}$ and $R_{B}=R_A$ of the 
species A and B, respectively). 

The corrugated wall containing pits with a square cross
section (see Sec. \Rnum{2}) has been placed at the
bottom of the computational box with the normal of the wall 
pointing into the $z-$direction. The radius of the wall particles
has been chosen as $R_C=R_{A}/3$. This choice of the 
size of the wall particles turns out to be a sound compromise 
between minimizing the effect of wall roughness on interfacial 
structures and the computational cost.
Periodic boundary conditions have been applied in the $x$ and $y$
directions. At the upper end of the box the boundary 
conditions prescribed for the number densities
$\rho_{A}$ and $\rho_{B}$ correspond to the liquid side of
liquid-vapor coexistence of the bulk of the binary liquid
mixture.
Alternatively to using the number densities, we also introduce the 
concentrations $c_A = \rho_{A}/(\rho_{A} + \rho_{B})$, 
$c_B = \rho_{B}/(\rho_{A} + \rho_{B})$, and the total fluid packing fraction
$\eta = \eta_{A} + \eta_{B}$ with $\eta_{A} = (4\pi/3)R^3_A\rho_A$,
$\eta_{B} = (4\pi/3)R^3_B\rho_B$; $\eta_{A}$ and $\eta_{B}$ 
are the fluid volume fractions blocked by the 
hard cores of the A and B fluid particles, respectively.
\onecolumngrid
\begin{center}
\begin{figure}[h]
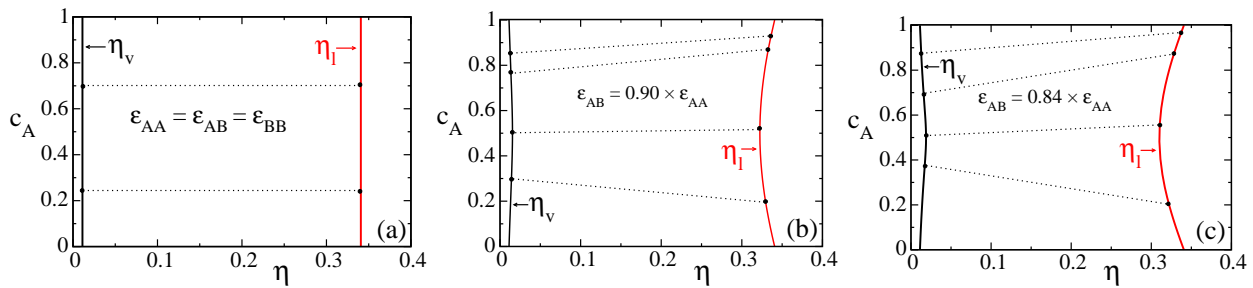
      
\vspace*{-0.0cm}
\hspace*{0.0cm}\includegraphics[scale = 0.20]{Fig2-a.eps}
\hspace*{0.0cm}\includegraphics[scale = 0.20]{Fig2-b.eps}
\hspace*{0.0cm}\includegraphics[scale = 0.20]{Fig2-c.eps}
{\vspace*{-0.00cm}
\begin{spacing}{0.0}
\caption{Phase diagrams of the binary liquid mixture 
(as discussed in the main text)
in the plane of total packing fraction $\eta$ and concentration $c_A$
of A particles ($c_B = 1 - c_A$)
for various ratios $\epsilon_{AB}/\epsilon_{AA}$  
($\epsilon_{AA} = \epsilon_{BB}$ in all cases).
The temperature is fixed such that $\epsilon_{AA}/(k_{B}T) = 0.9834$. 
The black and the red lines show the packing fractions at bulk coexistence
in the vapor and in the liquid phase, respectively, as a function of $c_A$.
Panel (a) corresponds to $\epsilon_{AB} = \epsilon_{AA} =\epsilon_{BB}$,
(b) to $\epsilon_{AB} = 0.90 \times \epsilon_{AA}$,
and (c) to $\epsilon_{AB} = 0.84\times \epsilon_{AA}$.
For $\epsilon_{AB}/\epsilon_{AA} \ne 1$, the concentrations 
$c_A$ and $c_B$ 
in the coexisting liquid (l) and vapor (v) phases, respectively,
are different. This is demonstrated by the tilt of the dotted 
tielines.} 
\end{spacing}}
\end{figure}
\end{center}


\begin{figure}[h]
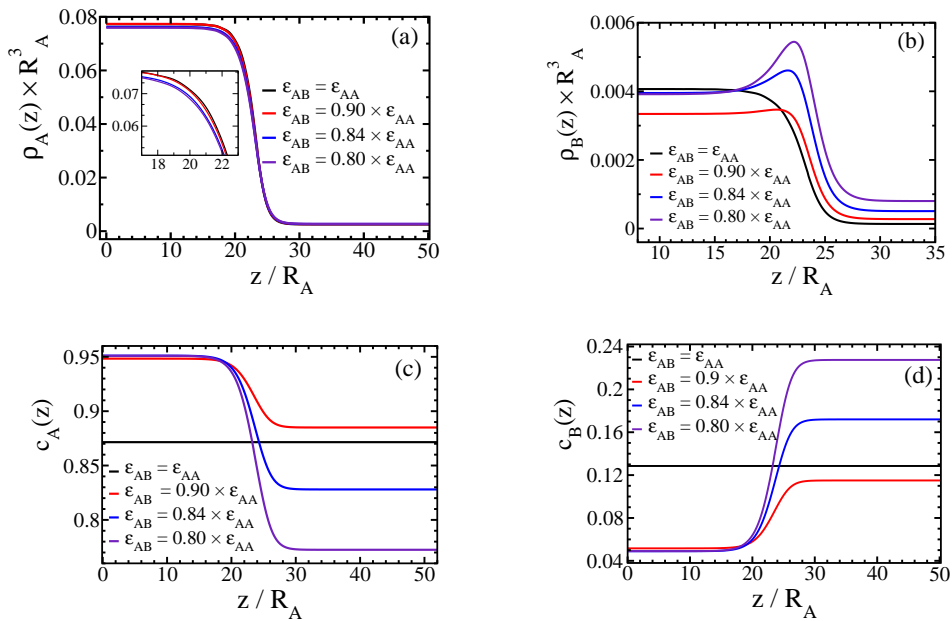
      
\hspace*{-0.0cm}\includegraphics[scale = 0.21]{Fig3-1.eps}
\hspace*{1.3cm}\includegraphics[scale = 0.19]{Fig3-2.eps}

\vspace*{0.54cm}

{\hspace*{0.5cm}\includegraphics[scale = 0.20]{Fig3-3.eps}
\hspace*{1.4cm}\includegraphics[scale = 0.20]{Fig3-4.eps}}
{\vspace*{-0.00cm}
 \begin{spacing}{0.0}
\caption{Number density and concentration profiles
for A and B particles across the free 
liquid-vapor interface for various values of 
$\epsilon_{AB}/\epsilon_{AA}$ and for the 
reduced temperature $\epsilon_{AA}/(k_{B}T) = 0.9834$.
The black line in panels (c) and (d), represents
the concentration profiles across the liquid - vapor 
interface. This is spatially constant for the binary 
liquid mixtures with $\epsilon_{AB} = \epsilon_{AA}$, which are
effectively one component fluids so that the {\it concentrations}
are the same in the vapor and liquid phases.} 
\end{spacing} }
\end{figure}
\twocolumngrid

In the following calculations at various concentrations
$c_{B}$ $(c_{A} = 1 - c_{B})$, always the concentrations in 
the bulk liquid adjacent to the corrugated wall have been 
controlled. The wall in the computational box contains a single pit.
The periodic boundary conditions imposed in the $x$ and $y$ directions imply, however,
that a wall with a periodic array of pits is considered.
But in the calculations presented here, the distance between
the pits in the array is sufficiently large so that
de facto liquid intrusion into isolated pits is studied. 

The strength of the interaction between like particles has been kept
fixed in all studies presented here, such that 
$\epsilon_{AA}/(k_{B}T) = 0.9834$ ($T$ is the absolute temperature
and $k_{B}$ is the Boltzmann constant). In most of our studies the pits
have a width of $w = 7\sigma$ and a depth of 
$D = 4\sigma$ ($\sigma = 2R_{A} = 2R_{B}$ is the diameter 
of the fluid particles), {\it i.e., } in the case of molecular fluids they 
are truly nanoscopic. Studying narrow and shallow pits reduces the
computational costs and opens also the opportunity to observe 
specific effects due to nanoconfinement.
\\ 
In a first series of 
computations, a fluid has been studied which effectively
behaves as a one-component fluid with respect to bulk properties.  
For this fluid model the fluid-fluid interaction strengths are all equal, {\it i.e.},
$\epsilon_{AA} = \epsilon_{AB} = \epsilon_{BB}$.
However, the A and the B particles are taken to interact with the solid wall with different 
strengths. The B particles are chosen to interact
more strongly with the wall than the A particles. Two 
different subvariants of this model have been studied in more
detail. In Sec. \Rnum{4}.A.1, studies are presented for which the 
interaction parameter $\epsilon_{A}$ of the A particles
with  the wall has been fixed to such a value that the planar 
wall is lyophobic for a pure A fluid, i.e., the macroscopic 
contact angle characterizing this fluid-wall system is 
$\approx 112^{\circ}$. The strength $\epsilon_{B}$ of the
interaction of the B particles with the wall has 
been fixed, too, such that $\epsilon_{B}/\epsilon_{A} = 2.33$.
By changing the concentration $c_{B}$ of the B particles in 
the fluid, the macroscopic contact angle $\theta_{Y}$ can be
tuned to values smaller than the one for the pure A fluid, due 
to the stronger interaction of the B particles with the wall. 
Thus, at bulk liquid-vapor coexistence, intrusion of the liquid into the pits can be 
studied as  function of
$c_{B}$. Furthermore, one can test the (meta)stability 
of the intruded Wenzel state versus the Cassie state,
in which the liquid remains above the pit and the pit is filled 
with its vapor. In order to test the metastability, the iterative
determination of the density distribution is typically initialized
with two different configurations. The first one is similar to the
Cassie configuration with bulk liquid densities above the pit 
and bulk vapor densities inside the pit with the
liquid-vapor interface placed close 
to the opening of the pit. The second initial configuration 
is one with the bulk liquid occupying the whole accessible
space.    

Another route to vary $\theta_{Y}$, within the special
fluid model chosen in the above first series of computations, consists of 
changing the interaction strengths $\epsilon_{A}$ and $\epsilon_{B}$
of the fluid-wall interaction. The results of these studies are 
presented in Sec. \Rnum{4}.A.2. In all these studies the outcome of the microscopic
DFT calculations concerning the intrusion is compared with the 
macroscopic prediction which, at bulk liquid-vapor coexistence,
simply states that for $\theta_{Y} < 90^{\circ}$ the liquid intrudes
the pit. Furthermore, the relative stability of the Wenzel and the 
Cassie states and their free-energy difference (provided that both states
do occur as (meta)stable states) are computed using both microscopic
DFT as well as the macroscopic theory. The input parameters
needed for the macroscopic theory are readily computed by 
carrying out independent calculations of the various interfacial
tensions at planar interfaces for the very same microscopic model
and the identical parameter set as used in the DFT studies for the
full problem of liquid intrusion. 

A key issue is, if, based on the value of
Young's contact angle alone, one can predict
as to whether a liquid intrudes a pit or not, or whether additional microscopic parameters,
not explicitly accounted for in a macroscopic theory, play a major role.
In order to shed additional light on this question we analyze the microscopic
details of the fluid structure and how these change as a 
result of variations of those parameters which are used to tune 
the contact angle $\theta_{Y}$. For instance, for a ratio 
$\epsilon_{B}/\epsilon_{A}$  substantially larger than
one, the concentration of B particles is considerably enhanced 
at the walls as compared to the bulk liquid. On the other hand, for 
the special fluid discussed above, which is a one-component 
fluid with respect to bulk properties, the concentration of B
particles is the same in the liquid and the vapor phase as well as across
a liquid-vapor interface, away from any wall. One could expect that 
such peculiarities, which are specific to a particular fluid/wall 
system, should be strongly relevant for determining the intrusion 
conditions, and that the values of pertinent parameters are needed in addition 
to the information encoded in Young's contact angle $\theta_{Y}$.  
This expectation is heightened by the fact that the surroundings at the 
opening of the pit, which should be of particular importance for the 
intrusion process, are very different from those implicitly assumed 
for the computation of $\theta_{Y}$. In narrow confinement, the role 
of microscopic details is expected to be even more pronounced. 

We have carried out further studies along this direction by conducting 
another series of computations in which we consider a bona fide binary liquid mixture. 
In this series, the interaction strength 
$\epsilon_{AB}$ between unlike fluid particles is taken to be different
from those between like particles $(\epsilon_{AA} = \epsilon_{BB})$.
The other parameters have the same or similar values as in the first series 
of computations discussed above. For various ratios 
$\epsilon_{AB}/\epsilon_{AA}$, the concentration $c_{B}$
has been varied, which leads to a variation of $\theta_{Y}$
(changing $\epsilon_{AB}/\epsilon_{AA}$ at fixed $c_{B}$
also leads to changes in $\theta_{Y}$). In the present studies,
ratios $\epsilon_{AB}/\epsilon_{AA} < 1$ have been 
considered. In this case, the concentration $c_{B}$ of the 
B particles in the vapor phase is enhanced relative to the
prescribed concentration of B particles in the liquid phase (see Fig. $2$). 
Accordingly, there is a gradient in $c_{B}$ across the liquid-vapor interface
with the higher values of $c_{B}$ occurring at the side pointing towards the pit
filled with vapor. This is distinct from the situation discussed above; 
the modified interplay of such a profile of $c_{B}$, with
$c_{B}$ enhanced at the walls, might be relevant for liquid intrusion.
For various ratios $\epsilon_{AB}/\epsilon_{AA}$, in Fig. $3$ we have displayed the 
corresponding density and concentration profiles for both
the A and the B particles, respectively, across the free liquid-vapor interface.
The results of these studies, in which the wall fluid interaction has 
been varied, are presented in Sec. \Rnum{4}.B.

Finally, in Sec. \Rnum{4}.C the results of certain studies are presented,
in which the effects are explored due to changing the width or the depth of the pit. 
These studies have been carried out for the model within 
which the fluid in the bulk is de facto a one-component one.

\subsection{Special binary fluid: one-component in the bulk, binary 
            with respect to the wall interactions}
In this subsection we discuss the wetting behavior of a special liquid,
which is a one-component liquid in the bulk ($R_{A} = R_{B}$, 
$\epsilon_{AA} = \epsilon_{AB} = \epsilon_{BB} $), but 
a binary one with respect to the interaction with the wall. 
The wall-liquid interactions are chosen such that the B 
particles are attracted more strongly by the wall than the A particles. 
We have varied the wall-fluid interaction strengths $\epsilon_{A}$
and $\epsilon_{B}$ as well as the ratio $\epsilon_{B}/\epsilon_{A}$. 
In the following, the wall - liquid interaction 
is represented in terms of the macroscopic Young contact angle 
$\theta_{Y}(c_{B} = 0)$ formed by a pure A type fluid on the 
corresponding planar wall. Because the properties of the pure A fluid are kept 
fixed in all calculations presented here, $\theta_{Y}(c_{B} = 0)$
uniquely characterizes the changes of the interaction between the
A particles and the wall. This  choice of representation has two benefits.
First, it facilitates intuition for the {\mbox{A - particle $-$}} 
wall interaction strength. Secondly, this representation of the interaction
strength is more robust with respect to numerical uncertainties 
resulting from descritizing the particle number densities on 
a grid which due to practical computational limitations 
cannot be refined as much that the values of
all quantities have fully converged. 
For instance, intrusion of a pure A liquid into a pit would 
occur at somewhat different values of $\epsilon_{A}$ for 
different mesh widths of the grid; however, it occurs 
at the same contact angle, calculated for the respective grid width.
First, in Sec. \Rnum{4}.A.1 we discuss the intrusion 
as a function of composition of the liquid, keeping $\epsilon_{A}$,
$\epsilon_{B}$, and $\epsilon_{B}/\epsilon_{A}$ fixed. 
In Sec. \Rnum{4}.A.2, we study the influence of 
the strengths of the fluid-solid interactions, keeping the composition 
of the liquid fixed. 

\subsubsection{Effect of composition}
In order to study how the composition of a binary liquid mixture 
affects its wetting behavior on a textured wall, we have
fixed $\epsilon_{AB} = \epsilon_{AA}(= \epsilon_{BB})$
and varied the composition of the liquid. We have started our analysis 
by considering a liquid which consists purely of A particles 
to which we gradually add B particles, until we observe intrusion.
The interaction strength $\epsilon_{A} = 0.9834 \times k_{B} T$
has been chosen such that the pure A liquid forms a contact angle 
$\theta_{Y}(c_{B} = 0) \approx 112^{\circ}$.
For the mixture with $c_{B} \ne 0$ a fixed ratio 
$\epsilon_{B}/\epsilon_{A} = 2.333$ has been used.
The composition in the liquid phase is controlled such that 
for each composition liquid-vapor coexistence is maintained.
As a prerequisite the macroscopic Young 
contact angle $\theta_{Y}$ on a corresponding
planar wall has been computed as a function of composition.
(In contrast to the contact angle measured from, 
{\it e.g.}, a spherical-cap shaped sessile drop of finite size,
the macroscopic Young contact angle $\theta_{Y}$ is independent
of how the location of the interface is defined. For a 
detailed discussion of this issue see Ref. \cite{Schim-nap-di}.) 
Increasing the concentration of B particles decreases the contact angle 
(see Fig. $4$) as a result of the stronger 
interaction of the B particles with the wall.
\begin{figure}[h]      
\vspace*{0.55cm}
\hspace*{0.0cm}\includegraphics[scale = 0.3]{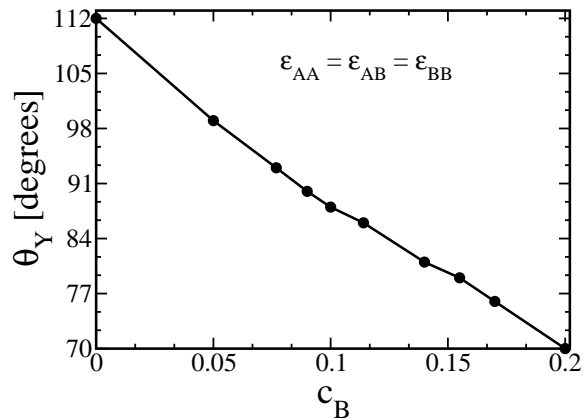}
{\vspace*{-0.0cm}
 \begin{spacing}{0.0}
\caption{Macroscopic contact angle $\theta_{Y}$ at a planar wall
as function of $c_{B}$ with fixed
$\epsilon_{AA} = \epsilon_{BB} = \epsilon_{AB}$. The
strength $\epsilon_{A} = 0.9834 \times k_{B}T$  
of the interaction between the wall and the A particles is chosen 
such that  Young's contact angle attains the value
$\theta_{Y} \approx 112^{\circ}$, with
$\epsilon_{B}/\epsilon_{A} = 2.333$ for $c_{B} = 0$. 
The full line is a guide to the eye 
through the DFT data points. The absolute values of the
deviations of $\theta_{Y}$ from
the converged values, for a given set of interaction 
parameters,  might be as high as $5^{\circ}$
for the grid on which the densities are discretized.
The estimate is based on computations for a one component
system. The relative quantity 
$\theta_{Y}(c_{B}) - \theta_{Y}(c_{B} = 0)$, 
with the above fixed value of $\theta_{Y}(c_{B} = 0)$, 
is more accurate by roughly 
one order of magnitude.}
 \end{spacing} }
\end{figure}

In the next step, liquid intrusion ({\it i.e.}, the Cassie $-$ Wenzel transition) 
into pits with square cross section is studied as function of the 
composition of the liquid. In order to detect the intrusion,
for various compositions we have calculated within DFT the 
density profiles of the binary liquid mixture, 
always at liquid-vapor coexistence. In order to observe the Cassie - Wenzel transition,
the iterative determination of the fluid density profiles is initialized
with a liquid-vapor interface placed at the opening of the pits. 
The pits are filled with the coexisting vapor, whereas the liquid
remains suspended on the top of the pits. 
We have determined that value of $\theta_{Y}$ below which 
intrusion occurs; according to the macroscopic theory
this is expected to occur for $\theta_{Y} < 90^{\circ}$.
We also carried out calculations with initializing the system in the  
Wenzel state (i.e., the pits are filled with liquid) and 
computed  the grand canonical potentials of both states, 
provided both are (meta)stable. These results are compared with 
the macroscopic predictions. At liquid-vapor coexistence, 
the difference between the grand canonical potentials of the 
Cassie-Baxter state $(\Omega_{CB})$ and of the Wenzel 
state $(\Omega_{W})$ can be computed within the 
macroscopic theory once the interfacial tensions are known; 
for a pit with square cross section of width $w$ 
and depth $D$ one has  
\be\label{eq:17}
\Omega_{CB} - \Omega_{W} = \Delta\Omega =
\sigma_{lv}w^{2} [1 + \frac{4D+w}{w} \cos\theta_{Y}].
\ee
Both states coexist, {\it i.e.}, $\Omega_{CB} = \Omega_{W}$, if 
\be\label{eq:18}
\cos\theta^{Y}_{c} = -\frac{w}{4D + w}
\ee
(see Eq. (\ref{eq:2}) in the Introduction). 

Here we discuss results for $w = 7 \times \sigma$
($\sigma = 2 \times R_{A}$), and $D = 4 \times \sigma$. For the pure
A liquid $(c_{B} = 0)$, the Cassie state is the stable state.
If the iteration is initialized in the Wenzel state, 
the latter turns 
out to be metastable with a higher grand canonical potential. If 
the liquid reaches $5\%$ of B particles $(c_{B} = 0.05)$, 
the  previously stable Cassie state becomes  metastable. 
The contact angle for this composition is $\theta_Y \approx 99^{\circ}$ 
(see Fig. $4$). For this composition the Wenzel state
has a lower grand canonical potential. The metastability of the Cassie state 
was tested by starting the iteration scheme with a liquid-vapor 
interface positioned one $\sigma$ deep inside the pit.
Once the iteration has converged, the liquid-vapor interface 
has moved up to the same position as the one obtained by using the 
standard Cassie state initialization in the case that the latter
is stable.  
For $c_{B} = 0.09$ the contact angle reduces to 
$\theta_{Y}\approx 90^{\circ}$ (see Fig. $4$), 
and the Cassie state still remains metastable.  
For $c_{B} \gtrsim 0.10$, $c_{B}$ has been ramped up in smaller 
steps of ca. $0.01$. The Cassie state remains metastable up to 
$c_{B} \approx 0.129$, corresponding to $\theta_{Y}\approx 83^{\circ}$.
Figure $5$ shows the density distributions for this liquid composition 
of $c_{B} \approx 0.129$. If the concentration of B 
particles is increased further up to $c_{B} \approx 0.138$, which corresponds
to $\theta_{Y} \approx 81^{\circ}$, the Cassie state becomes 
unstable. Both initial configurations (Cassie and Wenzel)
eventually converge into the Wenzel state (Fig. $6$). 
The number density of B particles is much higher 
at the walls as compared with that of the A particles.  
In Fig. $7$, we show the difference $\Delta \Omega$ of the 
grand canonical potentials of the Cassie and 
Wenzel states, for compositions 
between $c_{B} = 0$ and $c_{B}\approx 0.138$, as obtained from DFT. 
These  values are compared with 
$\Delta \Omega$ calculated from Eq. (\ref{eq:17}) for $w = 7\times \sigma$,
$D = 4 \times \sigma$ (red line), and for an alternative convention for defining 
the width and the depth of the pit, {\it i.e.}, 
$w = 6\times \sigma$, $D = 3\times \sigma$
(blue line). In the first convention the width is defined as the distance 
between the centers of the substrate particles in the topmost layers of the opposing wall 
surfaces, {\it i.e.}, including the depletion zones of the fluid 
densities; the depth is defined accordingly 
(see Fig. $1$(b)). 
In the second convention the width is
defined as the width of the space inside the pit which is actually
accessible to the fluid particles, thus the zones of 
strongly depleted fluid densities are not considered 
to be part of the pit volume. The reduced depth in the second convention
accounts for the fact that the liquid-vapor interface in the Cassie state
is not perfectly flat above the pit, but penetrates into the pit by 
ca. one $\sigma$ thus reducing the effective depth of the 
vapor filled space.  The contact angle $\theta^{Y}_{c}$ 
corresponding to $\Delta \Omega = 0$, as predicted by 
Eq. (\ref{eq:18}),
is $108^{\circ}$ or $109^{\circ}$, respectively, whereas 
$\theta^{Y}_{c}$ 
found from DFT is $109^{\circ}$. The macroscopic results for 
$\Delta \Omega$, based on two different conventions
for the geometric parameters width and depth,  differ considerably.
The second convention renders a better agreement
between the microscopic and the macroscopic results. 
However, none of the two acceptable alternative 
conventions for defining the geometric parameters leads to
a close agreement between the macroscopic
predictions for $\Delta \Omega$ and the microscopic results (see Fig. $7$). 

\subsubsection{Influence of the strength of the liquid-wall interaction}

In this section, we discuss the influence of the strength of the
liquid-wall interaction on the  wetting behavior of the liquid
and on the relative stability of the Cassie and Wenzel states.
We consider again a  particular liquid in the sense that it
is a one-component fluid with respect to its bulk properties.
The bulk composition of this liquid is fixed at $c_{B} = 0.05$,
but we vary the values of $\epsilon_{A}$, $\epsilon_{B}$,
and $\epsilon_{B}/\epsilon_{A}$. Here we discuss  the
results for $w = 7 \times \sigma$ and $D = 4 \times \sigma$.
We have considered two values of $\epsilon_{A}$, and for each
$\epsilon_{A}$ we have varied $\epsilon_{B}/\epsilon_{A}$.
First, we have taken the same value of $\epsilon_{A}$
as in the previous section,
which leads to $\theta_{Y} \approx 112^{\circ}$ for $c_{B} = 0$.
We have calculated the fluid number densities and the grand
canonical potential by starting the iteration
scheme both from Cassie and from Wenzel configurations.
For $\epsilon_{B}/\epsilon_{A} > 2.833$
({\it i.e.}, $\theta_{Y} < 83^{\circ}$), the only stable state
is the Wenzel state whereas the  Cassie state is unstable.
For $\epsilon_{B}/\epsilon_{A} = 2.833$
({\it i.e.}, $\theta_{Y} \approx 83^{\circ}$),
the Cassie state becomes metastable and the Wenzel
state is stable with a lower grand canonical potential.
We have decreased gradually the ratio
$\epsilon_{B}/\epsilon_{A}$, keeping $\epsilon_{A}$
fixed, and thus increased the contact angle gradually up to
$\approx 107^{\circ}$. The resulting values of
$\Delta \Omega$  are represented by the filled circles in Fig. $8$.
The Cassie state remains metastable up to the highest
contact angle $\theta_{Y} \approx 107^{\circ}$ investigated
in this series of computations.

\newpage
\onecolumngrid
\begin{center}
\begin{figure}    
\vspace*{0.0cm}
\hspace{0.0cm}\includegraphics[scale = 0.43]{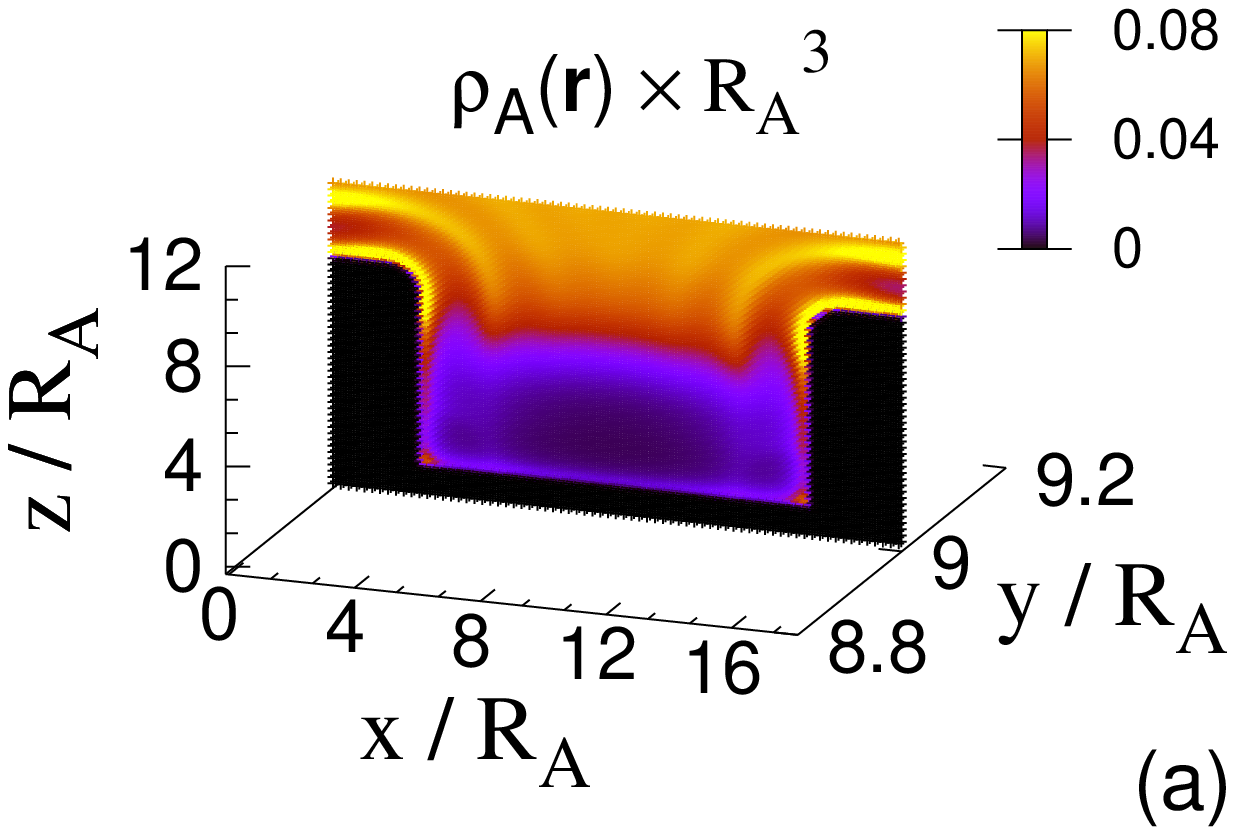}
\hspace*{1.8cm}\includegraphics[scale = 0.43]{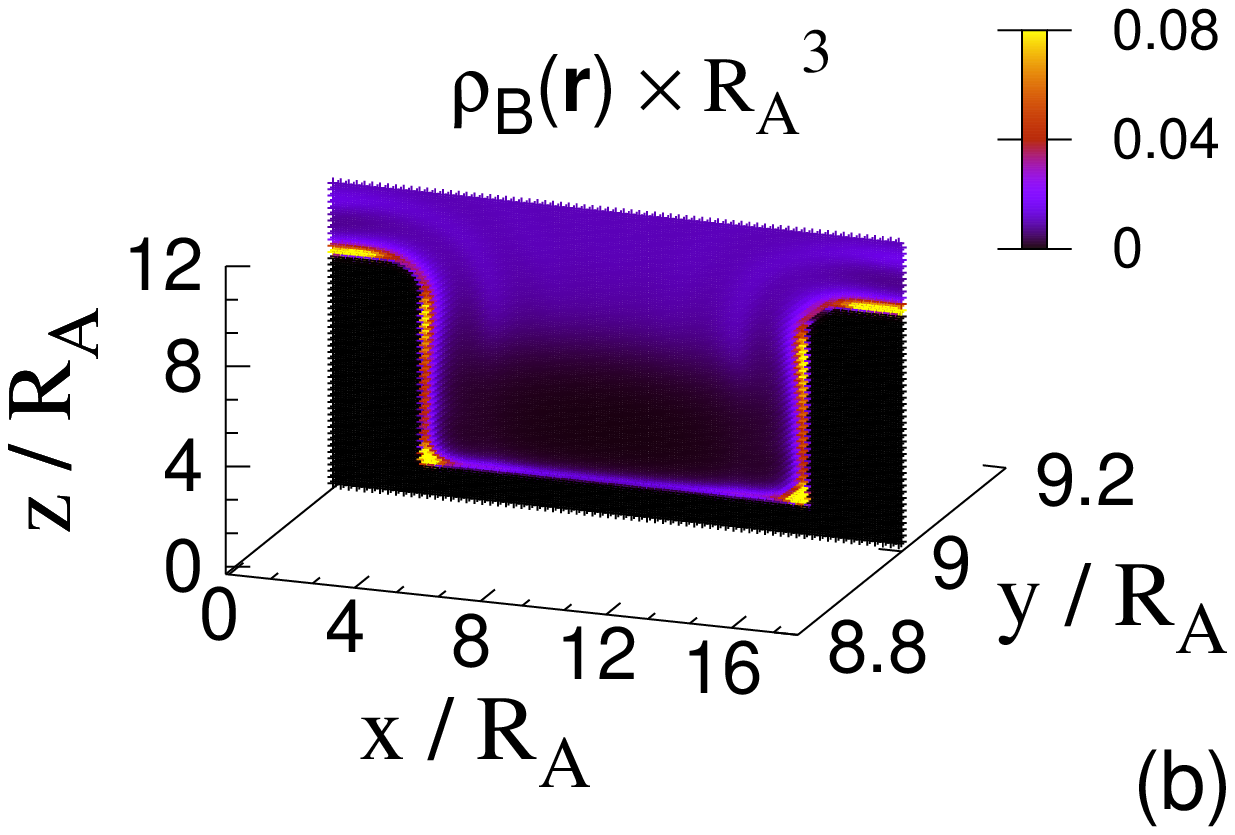}   

\vspace*{0.8cm}

\hspace*{-0.0cm}{\includegraphics[scale = 0.18]{Fig5-c.eps}}
\hspace*{1.8cm}{\includegraphics[scale = 0.43]{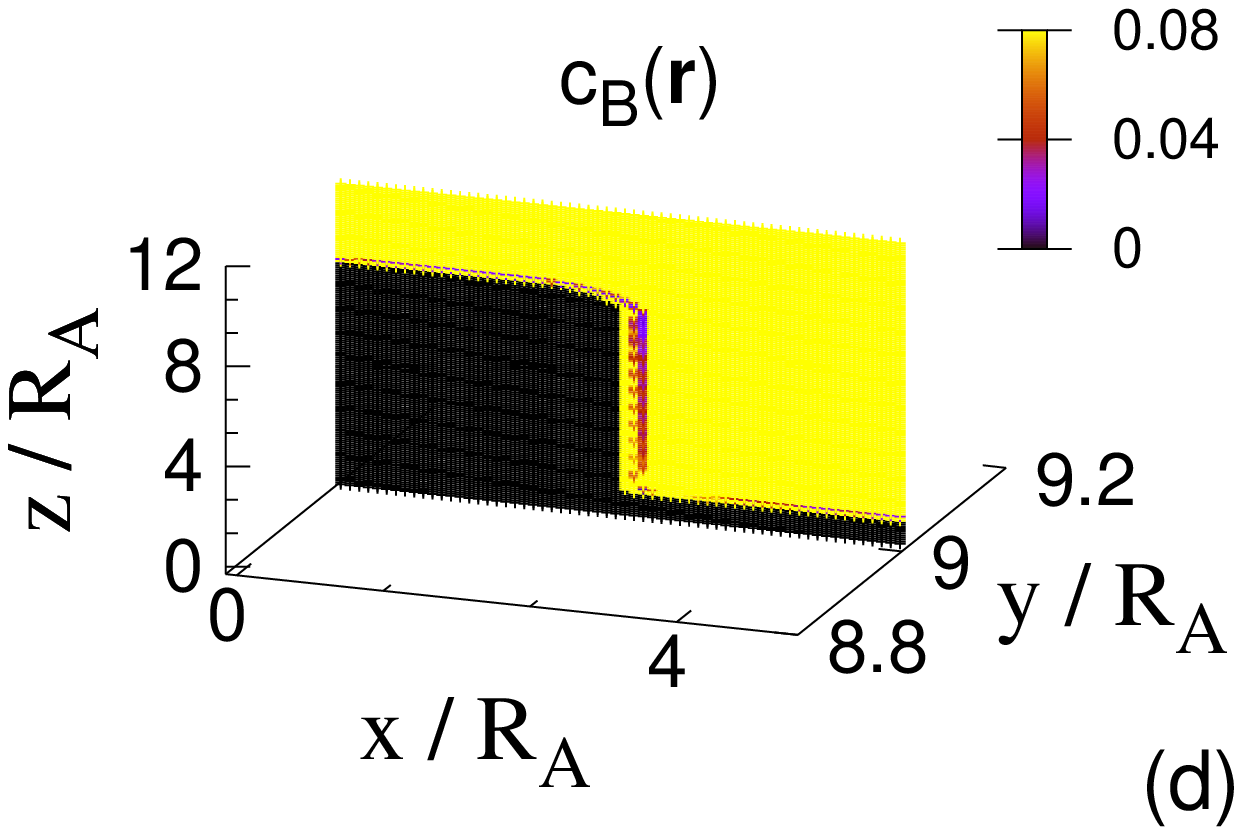}}
{\vspace*{-0.0cm}
 \begin{spacing}{0.0}
\caption{Number densities $\rho_{A}(\br)$  (a) and 
$\rho_{B}(\br)$  (b) for the bulk liquid with $c_{B} \approx 0.129$
and packing fraction $\eta_{l} = 0.3407$, in the $xz-$plane passing through the 
middle of the pit. With $\epsilon_{AA} = \epsilon_{BB} = \epsilon_{AB}$,
$\epsilon_{A}$ is chosen such that $\theta_{Y}(c_{B} = 0) \approx 112^{\circ}$
and with $\epsilon_{B}/\epsilon_{A} = 2.333$
for $c_{B} \ne 0$. For the concentration $c_{B}$ chosen here, one has 
$\theta_{Y} \approx 83^{\circ}$. The depth of the pit 
is $D = 4 \times \sigma$ and its width is $w = 7\times \sigma$
($\sigma = 2\times R_A$). 
The density distributions shown here correspond to a metastable
Cassie state. Panel (c) shows the densities along a line 
parallel to the $z$-axis and passing through the middle 
of the pit $(x = y = 9 \times R_{A})$. 
At contact wtih the wall 
the density of B particles is higher than
the density of A particles. There are 
some density oscillations in $\rho(z)$ at the opening, which
die out quickly. Panel (d) shows the 
concentration profile for B particles in the same plane 
as the one shown in panels (a) and (b), 
covering the left half of the pit and focusing  
on those parts of the pit where  potentially 
interesting phenomena, due to variations of $c_{B}$, occur.
}
 \end{spacing} }
\end{figure}
\end{center}

\begin{center}
\begin{figure}    
\vspace*{.50cm}
\hspace{0.0cm}\includegraphics[scale = 0.43]{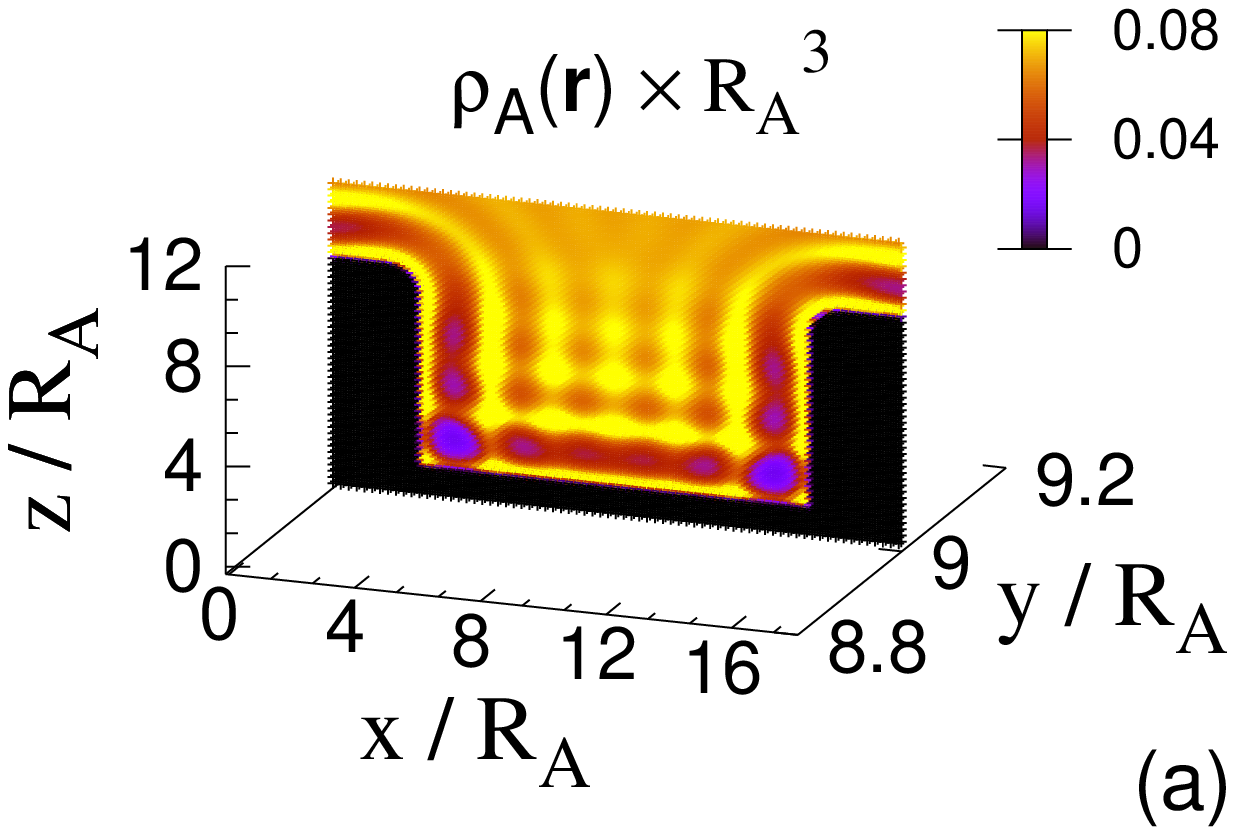}
\hspace*{0.4cm}\includegraphics[scale = 0.43]{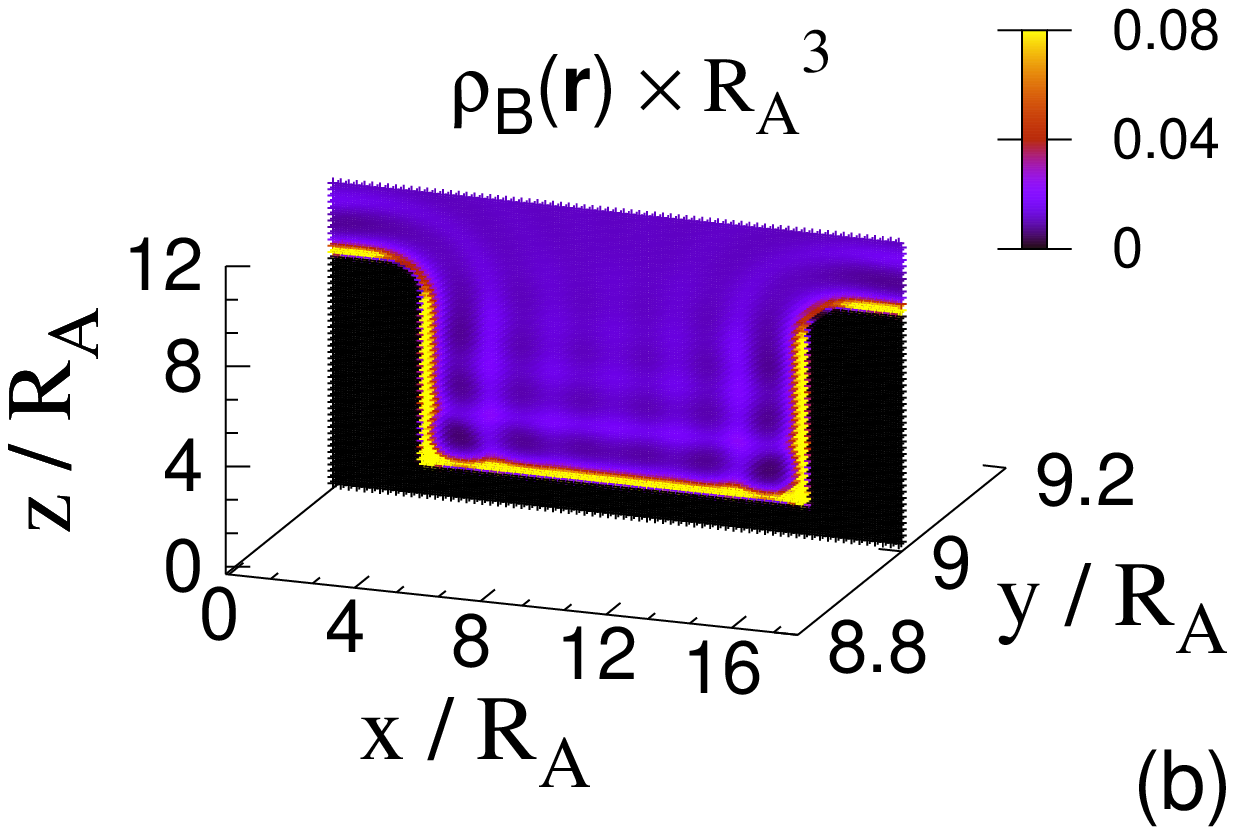}          
\hspace*{0.4cm}{\includegraphics[scale = 0.18]{Fig6-c.eps}}
{\vspace*{-0.0cm}
 \begin{spacing}{0.0}
\caption{Number densities $\rho_{A}(\br)$  (a) and 
$\rho_{B}(\br)$  (b) for the bulk liquid with $c_B = 0.1378$; the 
other parameters are the same as in Fig. $5$. 
The corresponding contact angle is 
$\theta_{Y} \approx 81^{\circ}$. The iteration
scheme has been initialized in the Cassie state, 
which turns out to be unstable for the chosen
set of parameters. Panel (c) shows the number 
densities along a line passing vertically through 
the center of the pit. 
}
\end{spacing} }
\end{figure}
\end{center}
\twocolumngrid

\begin{figure}[h]      
\vspace*{0.6cm}
\hspace*{0.0cm}\includegraphics[scale = 0.3]{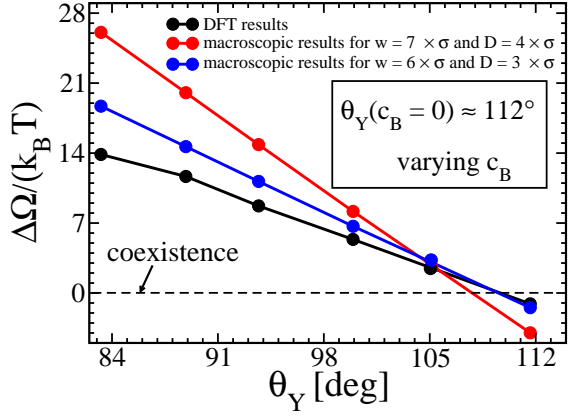}
{\vspace*{-0.0cm}
 \begin{spacing}{0.0}
\caption{Grand canonical free energy difference
$\Delta \Omega = \Omega_{CB} - \Omega_{W}$ 
for the Cassie-Baxter and Wenzel states as a 
function of $\theta_{Y}$. The contact angle
$\theta_{Y}$ is varied by changing $c_{B}$ at fixed  
$\epsilon_{AA}= \epsilon_{AB} = \epsilon_{BB}$, 
$w = 7 \times \sigma$, and $D = 4\times \sigma$. 
The value of $\epsilon_{A}$ has been chosen such that 
for $c_{B} = 0$ the contact angle is 
$\theta_{Y}(c_{B} = 0) \approx 112^{\circ}$. The black line 
corresponds to DFT results, the red line represents
the macroscopic predictions based on Eq. (\ref{eq:17}), 
with the pit width $w = 7 \times \sigma$ and depth 
$D = 4\times \sigma$. The blue line represents the 
macroscopic results for a pit with an effective 
width $w = 6\times \sigma$ and $D = 3\times \sigma$, 
taking into account the depletion zone and the slight 
intrusion of the liquid-vapor
interface into the pit even in the Cassie state.}
 \end{spacing} }
\end{figure}

In the next step we have
changed $\epsilon_{A}$ such that $\theta_{Y}$ 
associated  with a pure A liquid is ca. $124^{\circ}$. For this
value of $\epsilon_{A}$, we present results for 
the two ratios $\epsilon_{B}/\epsilon_{A} = 2.6$ and 
$\epsilon_{B}/\epsilon_{A} = 2.4$, indicated by  
stars in Fig. $8$. For $\epsilon_{B}/\epsilon_{A} = 2.6$, 
the corresponding contact angle is ca. $113^{\circ}$. 
For this value of $\theta_{Y}$, the Cassie state becomes 
stable whereas the Wenzel state becomes metastable. 
If $\epsilon_{B}/\epsilon_{A}$ is reduced to $2.4$, 
the contact angle increases to ca. $117^{\circ}$ 
with the Wenzel state still remaining metastable. When 
$\epsilon_{B}/\epsilon_{A}$ is reduced further, 
the Wenzel state becomes unstable and the only stable state is 
the Cassie state. 
  
We have compared these results of fully microscopic DFT 
computations with the results (data shown in black) 
obtained from Eq. (\ref{eq:17}).
The values of $\Delta \Omega$, obtained both microscopically and
macroscopically, are shown in Fig. $8$. The macroscopic calculations 
are carried out  for two sets of values for $w$ and $D$,
one for the actual pit dimensions ($w = 7\times\sigma$, 
$D = 4\times\sigma$; data shown in red) and one for the effective 
pit dimensions ($w = 6\times\sigma$, $D = 3\times\sigma$; 
data shown in blue).
Using the effective pit dimensions improves the agreement with 
the DFT results, but there are still considerable discrepancies.
The microscopic computations show that the Cassie and Wenzel 
states coexist at $\theta_{Y} \approx 110^{\circ}$.
Macroscopically, the coexistence of the Cassie and the 
Wenzel states for the actual ($w = 7\times\sigma$, 
$D = 4\times\sigma$) and for the effective pit dimensions 
($w = 6\times\sigma$, $D = 3\times\sigma$) occurs at  
$\theta_{Y}\approx 108^{\circ}$ and $\theta_{Y}\approx 109^{\circ}$,
respectively. 
\begin{figure}[h]      
\vspace*{0.6cm}
\hspace*{0.0cm}\includegraphics[scale = 0.3]{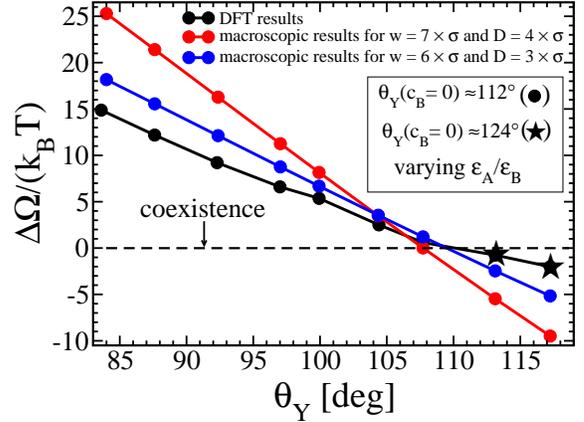}
{\vspace*{-0.0cm}
 \begin{spacing}{0.0}
\caption{Grand canonical free energy difference
$\Delta \Omega = \Omega_{CB} - \Omega_{W}$ 
for the Cassie-Baxter and Wenzel states as function 
of $\theta_{Y}$. The macroscopic 
contact angle $\theta_{Y}$ has been varied  
by changing $\epsilon_{A}$ and $\epsilon_{B}/\epsilon_{A}$. 
The black line shows DFT results, whereas the red 
and the blue lines correspond to the macroscopic 
predictions for the actual and the effective values of
$w$ and $D$, respectively.  We have chosen two values 
of $\epsilon_{A}$ resulting in 
$\theta_{Y} \approx 112^{\circ}$ (shown by filled circles) 
and $\theta_{Y} \approx 124^{\circ}$ (shown by stars)
for a pure A liquid. For each of these two 
values of $\epsilon_{A}$, $\epsilon_{B}/\epsilon_{A}$ 
is varied in order to change $\theta_{Y}$. 
The other parameters are fixed at $c_{B} = 0.05$, 
$\epsilon_{AA} = \epsilon_{AB} = \epsilon_{BB}$, 
$w = 7\times \sigma$, and $D = 4\times \sigma$.}
\end{spacing} }
\end{figure}

\begin{figure}[h]      
\vspace*{0.50cm}
\hspace*{0.0cm}\includegraphics[scale = 0.30]{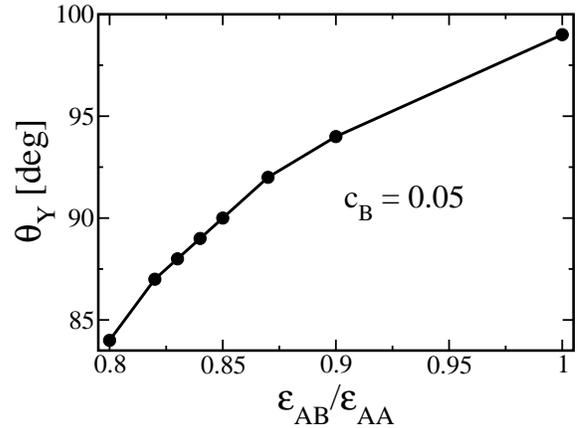}
{\vspace*{-0.0cm}
 \begin{spacing}{0.0}
\caption{Contact angle $\theta_{Y}$ as a function 
$\epsilon_{AB}/\epsilon_{AA}$ for $c_{B} \approx 0.05$
and $\epsilon_{AA} = \epsilon_{BB}$. The parameter 
$\epsilon_{A} = 0.9834 \times k_{B} T$ is chosen such that 
$\theta_{Y} \approx 112^{\circ}$
for $c_{B} = 0$, and $\epsilon_{b}/\epsilon_{A} = 2.333$.
The line is a guide to the eyes for the DFT data points.}
 \end{spacing} }
\end{figure}
\subsection{Binary liquid}
In this subsection, we study the wetting behavior as well as the 
metastability of the Cassie and Wenzel states for a liquid which is 
a bona fide binary liquid mixture also with respect to its 
bulk properties; but as a function of various parameters 
it is taken to be symmetric upon interchanging
$A$ and $B$ ($R_{A} = R_{B}$, $\epsilon_{BB} = \epsilon_{AA}$,
but $\epsilon_{AB} \neq \epsilon_{AA}$). In Sec. \Rnum{4}.B.1, 
we discuss the effect of varying $\epsilon_{AB}$, 
keeping the liquid-wall interaction and the
composition of the liquid fixed.
In the subsequent Sec. \Rnum{4}.B.2, we study the intrusion of the 
liquid as a function of $c_{B}$ for various values of 
$\epsilon_{AB}$ at fixed liquid-wall interaction. 
In Sec. \Rnum{4}.B.3, we analyze the influence of 
the liquid-wall interaction on the wetting behavior of 
the binary liquid mixture.

\subsubsection{Influence of $\epsilon_{AB}$}

In order to study the influence of $\epsilon_{AB}$ on the wetting behavior, 
we have fixed the composition of the liquid at $c_{B} \approx 0.05$,
the A $-$ A  and B $-$ B interaction strengths at 
$\epsilon_{AA}/(k_{B}T) = \epsilon_{BB}/(k_{B}T) = 0.9834$,
and decreased the ratio $\epsilon_{AB}/\epsilon_{AA}$ from $1.0$,
in steps of $0.05$, to $\epsilon_{AB}/\epsilon_{AA} = 0.90$,
and then from $\epsilon_{AB}/\epsilon_{AA} = 0.90$, 
in smaller steps of $0.02$, to $\epsilon_{AB}/\epsilon_{AA} = 0.80$.
For $\epsilon_{AB} = 0.90 \times \epsilon_{AA}$ and 
$\epsilon_{AB} = 0.84 \times \epsilon_{AA}$, the phase diagram 
of the liquid as a function of $c_{A} = 1 - c_{B}$, is shown in Fig. $2$.
For $\epsilon_{AB}/\epsilon_{AA}\neq 1.0$, the composition of
the coexisting vapor phase is not the same as that of the 
liquid phase, {\it i.e.}, the tielines are not 
horizontal (Fig. $2$).
The fluid-wall interaction strengths 
$\epsilon_{A}$ and $\epsilon_{B}$ are kept to be the same as 
in Sec. \Rnum{4}.A.1. As a prerequisite, Young's 
contact angle $\theta_{Y}$ on a planar wall is determined as a 
function of $\epsilon_{AB}$. In Fig. $9$, $\theta_{Y}$ is shown as a 
function of $\epsilon_{AB}$ for fixed composition $c_{B} \approx 0.05$. 
Upon lowering $\epsilon_{AB}/\epsilon_{AA}$, the contact angle 
decreases from $\theta_{Y} \approx 99^{\circ}$
at $\epsilon_{AB} = \epsilon_{AA}$ to $\theta_{Y} \approx 84^{\circ}$
for $\epsilon_{AB} = 0.80 \times \epsilon_{AA}$.

Next, we discuss the conditions for intrusion of the liquid into 
the pits with $w = 7 \times \sigma$ and $D = 4\times \sigma$. 
From the studies presented in Sec. \Rnum{4}.A.1, 
it is known that for $c_{B} = 0.05$ and $\epsilon_{AB} = \epsilon_{AA}$
the Cassie state is metastable and the Wenzel state is stable.  
If $\epsilon_{AB}$ is decreased, the Cassie state remains metastable 
up to $\epsilon_{AB} = 0.86 \times \epsilon_{AA}$ (corresponding to
$\theta_{Y} \approx 91^{\circ}$). For 
$\epsilon_{AB} = 0.84 \times \epsilon_{AA}$
(corresponding to $\theta_{Y} \approx 89^{\circ}$), the Cassie state 
turns unstable and the Wenzel state becomes the only stable state, 
to which the iteration procedure converges for any initial condition.
In Fig. $10$, we compare the grand canonical potential difference
$\Delta \Omega = \Omega_{CB} - \Omega_{W}$ obtained within 
DFT, with $\Delta \Omega$ calculated from Eq. (\ref{eq:17}) for 
various values of $\epsilon_{AB}$, expressed in terms 
of the corresponding contact angle $\theta_{Y}$. 
\\

\subsubsection{Intrusion as a function of the concentration
$c_{B}$ for various interaction strengths $\epsilon_{AB}$}

We have also studied the intrusion and the metastability of 
the Cassie and the Wenzel states for various values of 
$\epsilon_{AB}$ as a function of $c_{B}$, for each $\epsilon_{AB}$. 
For  $\epsilon_{AB} = 0.90 \times \epsilon_{AA}$, 
$w = 7 \times \sigma$, and  $D = 4\times \sigma$ 
the Cassie state is  metastable for $c_{B} \approx 0.05$ 
whereas the Wenzel state is the stable state 
(Sec. \Rnum{4}.B.1). 
We have increased $c_{B}$ from $ca.$ $0.05$ onwards 
in steps of $\approx 0.01$. The Cassie state remains metastable up to 
$c_{B} \approx 0.075$, for which the Wenzel state is 
the stable one. The corresponding 
Young's contact angle is $ca.$ $86^{\circ}$.                
These values correspond to the results shown in
Fig. $11$. Upon increasing $c_{B}$ beyond
$0.085$ (Fig. $12$),
thereby decreasing $\theta_{Y}$ to $84^{\circ}$, the Cassie
state becomes unstable. 
\begin{figure}[h]   
\vspace*{0.5cm}
\hspace*{0.0cm}\includegraphics[scale = 0.30]{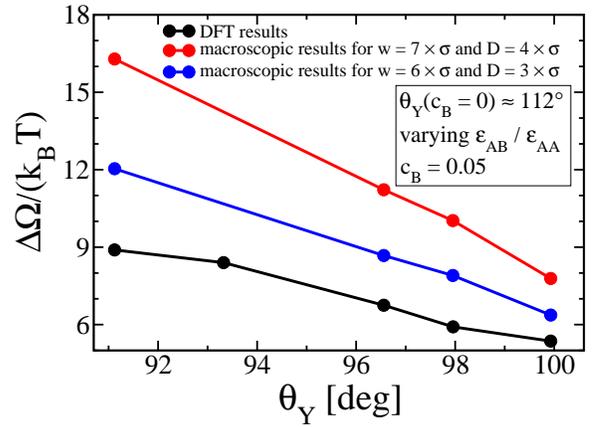}
{\vspace*{-0.0cm}
 \begin{spacing}{0.0}
\caption{Grand canonical free energy difference 
$\Delta \Omega =  \Omega_{CB} - \Omega_{W}$ for the
Cassie-Baxter and Wenzel states as function of 
$\epsilon_{AB}$ (expressed in terms of $\theta_{Y}$,
see Fig. $9$), for $\epsilon_{AA}= \epsilon_{BB}$, 
and $c_{B} = 0.05$. The black line interpolates the DFT results, 
whereas the red and blue lines represent 
$\Delta \Omega$ as obtained from the macroscopic theory 
for the actual parameter values $w = 7 \times \sigma$, 
$D = 4\times \sigma$ and the effective ones 
$w = 6\times \sigma$, $D = 3\times \sigma$,
respectively. The interaction strength 
$\epsilon_{A} = 0.9834 \times k_{B}T$ 
of the A particles with the wall is chosen such that the contact angle 
formed by the pure A liquid on the corresponding planar wall
is $\theta_{Y}(c_{B} = 0) \approx 112^{\circ}$.
The relative strength of the wall interactions is 
fixed at $\epsilon_{B}/\epsilon_{A} = 2.333$.}
 \end{spacing} }
\end{figure}

Upon decreasing $\epsilon_{AB}$
to $0.87 \times \epsilon_{AA}$ and keeping all other parameters fixed,
the liquid intrudes the pits at a lower concentration
$c_{B} \approx 0.064$ but at a higher contact angle $\theta_{Y}$
(see Fig. $13$). For $\epsilon_{AB} = 0.84 \times \epsilon_{AA}$,
the Cassie state remains unstable for concentrations of B
particles higher than $c_{B} = 0.037$; at the next
smaller concentration tested, $c_{B}\approx 0.025$,
it becomes metastable. The corresponding contact angles
are $\theta_{Y}\approx 94^{\circ}$ for $c_{B} \approx 0.037$
and $\theta_{Y} \approx 96^{\circ}$ for $c_{B} \approx 0.025$.
In Fig. $13$, we have introduced the
critical angle $\theta_{i}$ for intrusion as the mean of the
largest contact angle $\theta_{Y}$ below which intrusion
is observed and the smallest angle $\theta_{Y}$ at and
above which the Cassie state is found to be metastable.
This gap amounts to $ca.$ $2^{\circ}$ in the present studies
(the steps in which $\theta_{Y}$ is varied)
and defines the error bars indicated in Fig. $13$.
Obviously the critical angle $\theta_{i}$ of intrusion
increases upon decreasing the ratio
$\epsilon_{AB}/\epsilon_{AA}$ and is not fixed at $90^{\circ}$
as predicted by the macroscopic theory.

\onecolumngrid
\begin{center}
\begin{figure}[h]    
\vspace*{-0.0cm}
\hspace{-0.0cm}\includegraphics[scale = 0.43]{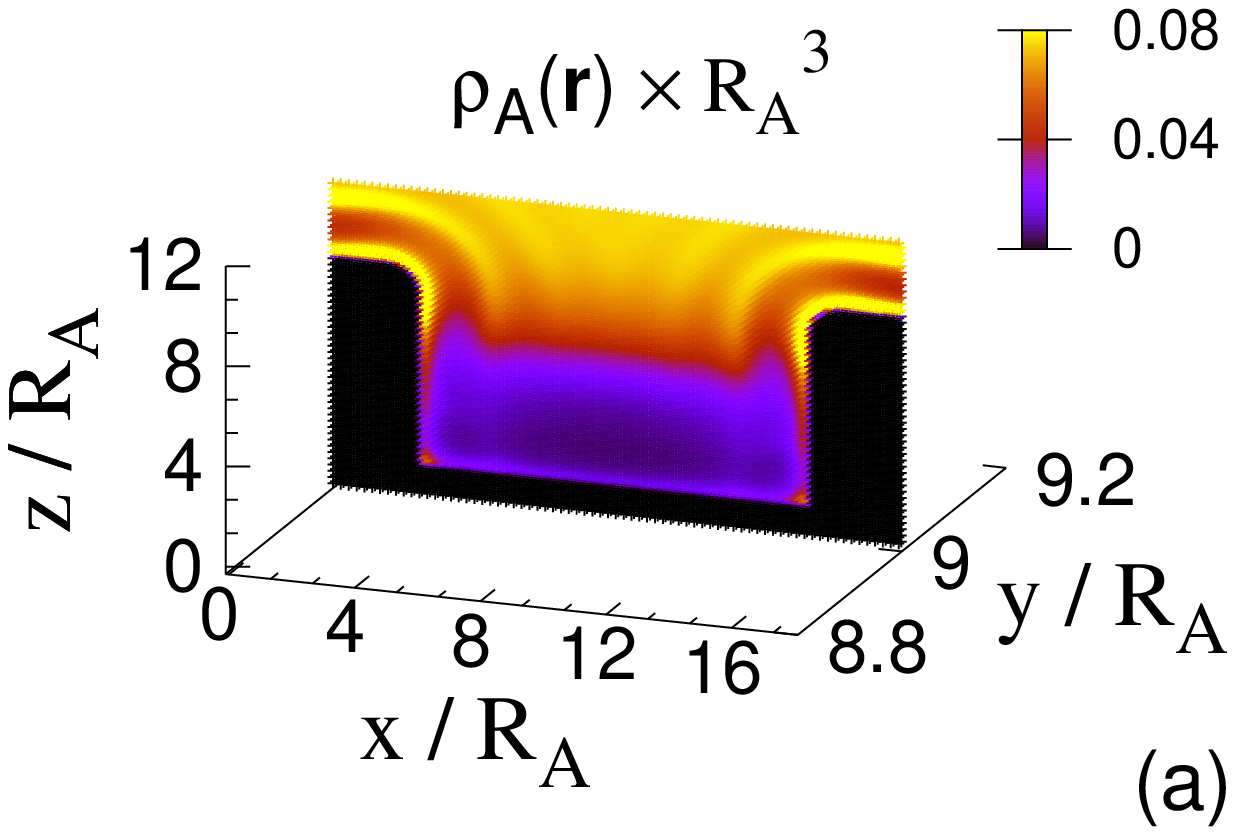}
\hspace*{1.8cm}\includegraphics[scale = 0.43]{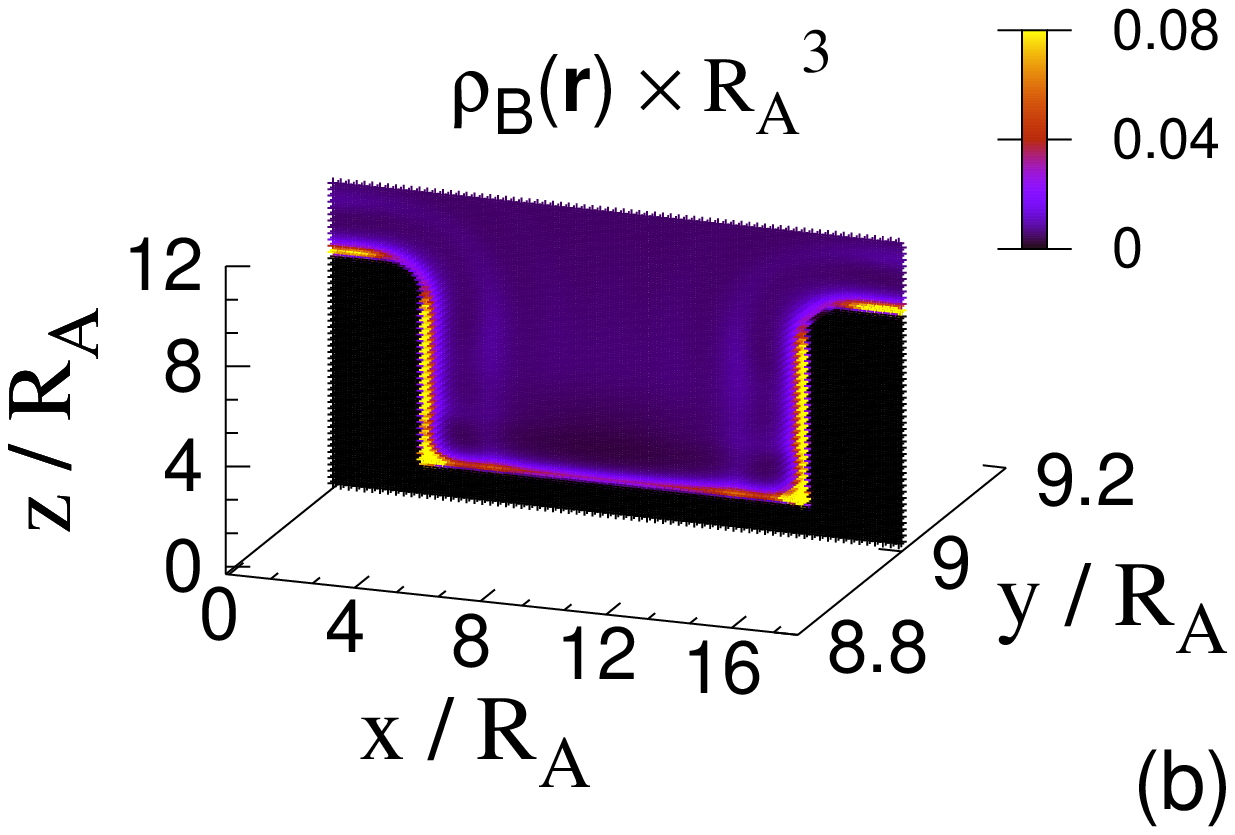}          

\vspace*{1.0cm}

\hspace*{-0.0cm}{\vspace*{-0.0cm}\includegraphics[scale = 0.18]{Fig11-c.eps}}
\hspace*{1.8cm}{\includegraphics[scale = 0.43]{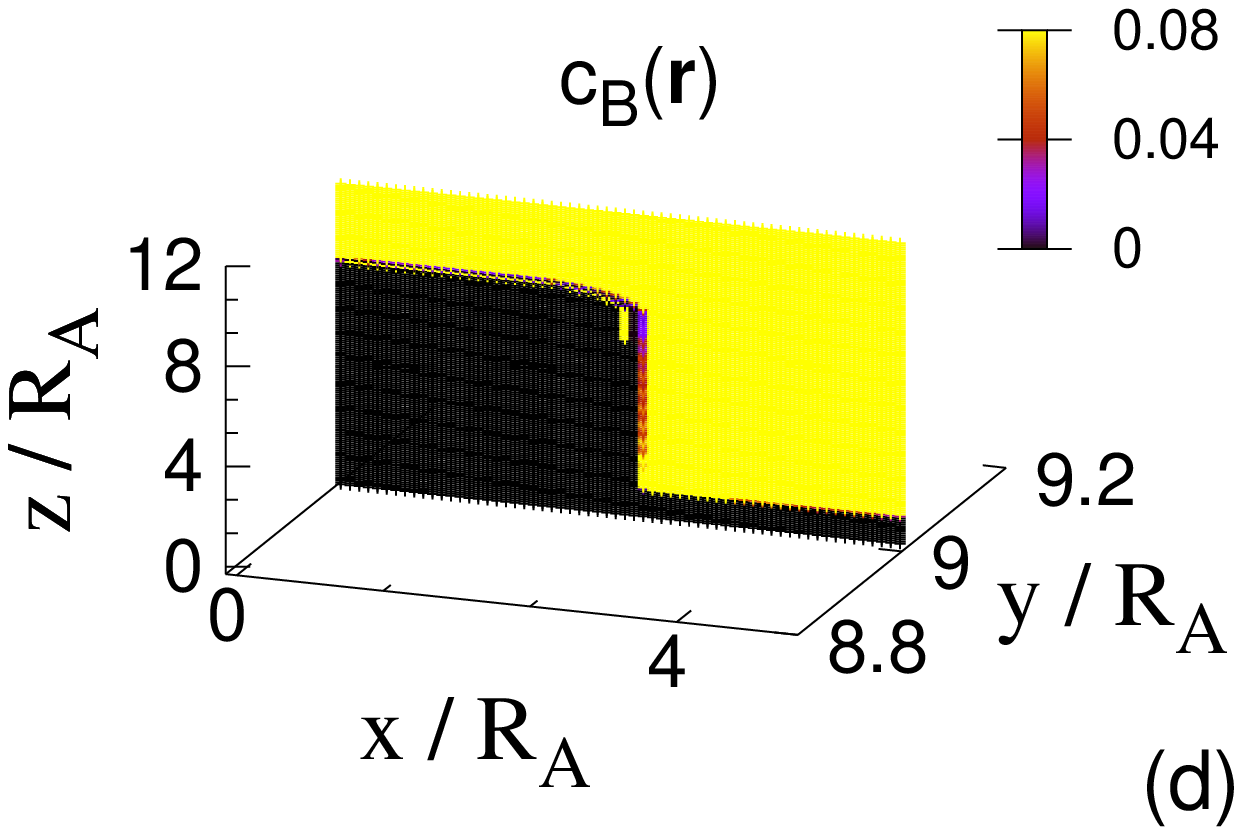}}

{\vspace*{-0.0cm}
 \begin{spacing}{0.0}
\caption{Number densities $\rho_{A}(\br)$  (a) and
$\rho_{B}(\br)$  (b) for the liquid with $c_B 
\approx 0.075$ and $\epsilon_{AB} = 0.9 \times \epsilon_{AA}$, 
in the $xz-$plane passing through the center of the pit. 
Panel (c) shows the number density along a line parallel 
to the $z$-axis and passing through the center of the pit. 
For the given composition of the liquid phase
the bulk liquid exhibits the packing fraction $\eta_{l} = 0.3357660$, 
whereas the coexisting vapor has $\eta_{v} = 0.0124869$. 
For this composition of the liquid bulk phase one has 
$\theta_{Y}\approx 86^{\circ}$.
Panel (d) shows the concentration profile for 
the B particles in the same plane as shown in 
panels (a) and (b), focusing
on those parts of the pit in which potentially 
interesting variations of $c_{B}$ occur. 
The fluid - wall interaction strengths and the pit dimensions
are the ones used in Fig. $10$.}
 \end{spacing} }
\end{figure}
\end{center}

\begin{center}
\begin{figure}[h]    
\vspace*{0.0cm}
\hspace{-0.0cm}\includegraphics[scale = 0.43]{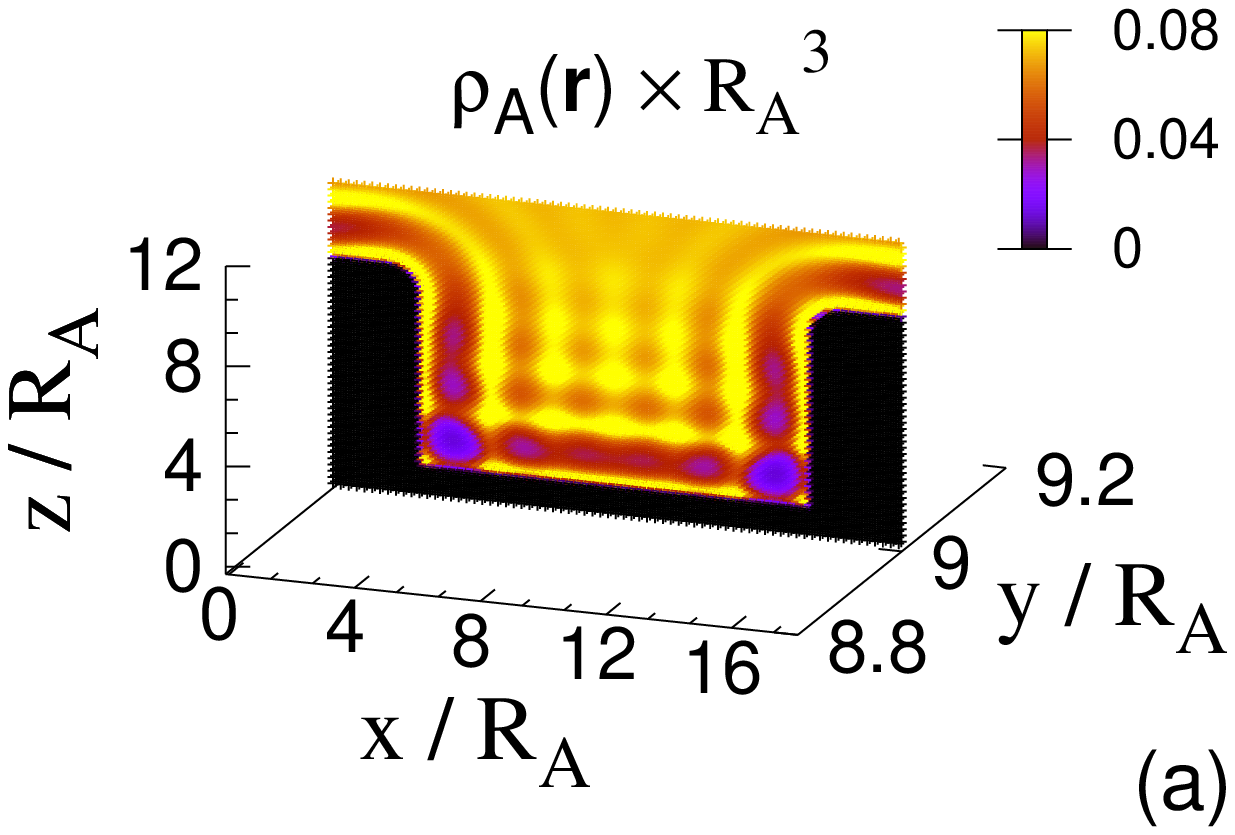}
\hspace*{0.4cm}\includegraphics[scale = 0.43]{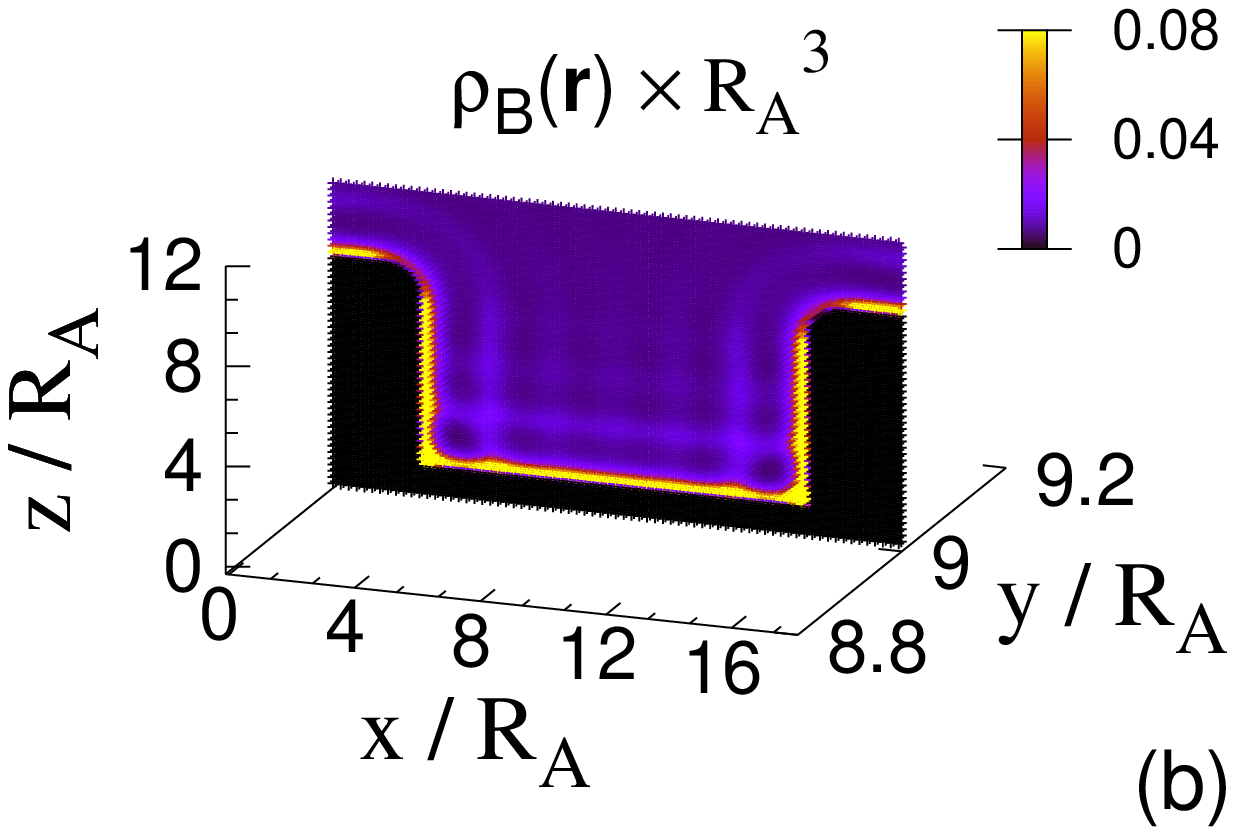}        
\hspace*{0.4cm}{\includegraphics[scale = 0.18]{Fig12-c.eps}}
{\vspace*{-0.0cm}
 \begin{spacing}{0.0}
\caption{The same as in Fig. $11$ for  $c_B \approx 0.085$
in the bulk liquid phase. Packing fractions of the coexisting
liquid and vapor phases in the bulk are 
$\eta_{l} = 0.3352940$ and $\eta_{v} = 0.0128612$,
respectively. The concentration of B particles 
in the bulk vapor phase is $c_{B} \approx 0.173$. The 
contact angle for the present system is 
 $\theta_{Y} \approx 84^{\circ}$.}
 \end{spacing} }
\end{figure}
\end{center}
\twocolumngrid
\twocolumngrid
\begin{figure}[]      
\vspace*{0.0cm}
\hspace*{0.0cm}\includegraphics[scale = 0.3]{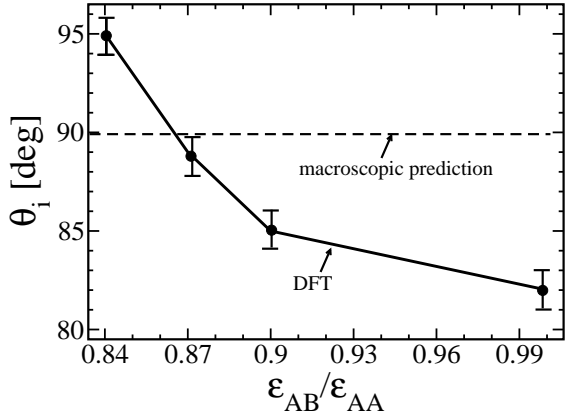}
{\vspace*{-0.0cm}
 \begin{spacing}{0.0}
\caption{ Critical contact angle $\theta_{i}$, below which the Cassie state
becomes unstable and liquid intrudes into the 
pits for four ratios $\epsilon_{AB}/\epsilon_{AA}$.
For each ratio the contact angle is varied by changing
the concentration $c_{B}$ such that intrusion occurs 
at $\theta_{Y} = \theta_{i}$. The strength $\epsilon_{A}$ is chosen such that
the contact angle formed by the pure A liquid is $112^{\circ}$
and $\epsilon_{B}/\epsilon_{A} = 2.333$. The
pit width is $w = 7\times \sigma$ and the depth is 
$D = 4 \times \sigma$.
The critical contact angles $\theta_{i} \approx 82^{\circ},
85^{\circ}, 89^{\circ},$ and $95^{\circ}$ 
correspond to $c_{B} = 1.0, 0.90, 0.87,$ and $0.84$, 
respectively.} 
 \end{spacing} }
\end{figure}
\subsubsection{Influence of the strength of the liquid-wall interaction}

Here the studies reported in Sec. \Rnum{4}.A.2 concerning
the influence of the strength of the liquid-wall
interaction on the relative stability of the Cassie and the
Wenzel states are extended  to ratios
$\epsilon_{AB}/\epsilon_{AA} \neq 1$.
For all the analyses in the  present section, the pit dimensions are
kept fixed at $w = 7 \times \sigma$ and $D = 4 \times \sigma$,
and the concentration of B particles in the bulk
is always taken to be  $c_{B} \approx 0.05$.
Two ratios $\epsilon_{AB}/\epsilon_{AA}$ ($0.9$, Fig. $14$)
and ($0.84$, Fig. $15$) are
considered and in each case the fluid-wall interaction strength
$\epsilon_{A}$ and the ratio $\epsilon_{B}/\epsilon_{A}$
of the two fluid-wall interaction strengths have been varied.
We start with discussing the case of
$\epsilon_{AB}/\epsilon_{AA} = 0.90$ and
consider three values of $\epsilon_{A}$ and various ratios
$\epsilon_{B}/\epsilon_{A}$ for each $\epsilon_{A}$.

First, we have chosen $\epsilon_{A} = 0.9834 \times k_{B}T$,
which was considered in the previous section, resulting
in a contact angle $\theta_{Y}(c_{B} = 0) \approx 112^{\circ}$ for a
pure A liquid. For $\epsilon_{B}/\epsilon_{A} = 2.833$,
the Wenzel state is the stable state whereas the Cassie
state is unstable and thus 
$\Delta \Omega = \Omega_{CB} - \Omega_{W}$
is undefined. Therefore, in Fig. $14$ no data point 
appears at the corresponding  contact angle
$\theta_{Y}(c_{B} = 0.05) \approx 84^{\circ}$. 
If $\epsilon_{B}/\epsilon_{A}$
is decreased to $2.766$, so that $\theta_{Y}$ increases to
$\theta_{Y}(c_{B} = 0.05) \approx 87^{\circ}$, the Cassie
state becomes metastable whereas the Wenzel state remains the 
stable state $(\Delta \Omega > 0)$.
Upon decreasing the ratio $\epsilon_{B}/\epsilon_{A}$ further to
$2.333$, the Cassie state remains metastable (as reported in the previous
section) and the Wenzel state remains stable, exhibiting
the lower grand canonical potential $(\Delta \Omega > 0)$. 
The resulting differences
$\Delta \Omega = \Omega_{CB} - \Omega_{W}$ between the
grand canonical potentials of the Cassie-Baxter and the
Wenzel states (corresponding to the aforementioned
first choice of $\epsilon_{A}$ and to the various
values of $\epsilon_{B}/\epsilon_{A}$
considered) are shown by filled circles in Fig. $14$.
In the second step, $\epsilon_{A} = 0.8195 \times k_{B}T$ 
is decreased such that the contact angle formed
by the pure A liquid on the corresponding planar
wall increases to $\theta_{Y}(c_{B} = 0) \approx 124^{\circ}$.
The resulting values of $\Delta \Omega$ are shown by
stars in Fig. $14$. For $\epsilon_{B}/\epsilon_{A} = 2.6$,
with a corresponding contact angle 
$\theta_{Y}(c_{B} = 0.05) \approx 110^{\circ}$,
the Cassie state still remains metastable and the Wenzel state is
the stable state ($\Delta \Omega > 0$). For 
$\epsilon_{B}/\epsilon_{A} = 2.4$, so that 
$\theta_{Y}(c_{B} = 0.05) \approx 113^{\circ}$, the Cassie and the Wenzel
states coexist ($\Delta \Omega = 0$). In the previous studies for a fluid with
$\epsilon_{AB} = \epsilon_{AA}$ and various fluid-wall interaction
strengths ($\epsilon_{A}$, $\epsilon_{B}/\epsilon_{A}$),
and for the same pit dimensions, the coexistence of the
Cassie and the Wenzel states has been detected at
$\theta_{Y}(c_{B} = 0.05) \approx 110^{\circ}$ (see Fig. $8$).
This implies that the contact angle $\theta^{Y}_{c}$, at which the
two states coexist, has changed by $3^{\circ}$ as a result of
changing certain microscopic details of the fluid-wall system whereas
the macroscopic theory (Eq. (\ref{eq:18})) predicts that 
$\theta^{Y}_{c}$ depends
only on the geometric parameters $w$ and $D$ and not on any other
details.

In the third step,  $\epsilon_{A} = 0.6556 \times k_{B}T$ 
has been decreased further, such that the corresponding 
contact angle for a pure A liquid increases to 
$\theta_{Y}(c_{B} = 0) \approx 137^{\circ}$. 
For the ratio $\epsilon_{B}/\epsilon_{A} = 3.25$, which corresponds
to $\theta_{Y}(c_{B} = 0.05) \approx 123^{\circ}$ 
for the mixture, the Wenzel state still remains metastable and the Cassie
state stable ($\Delta \Omega < 0$). 
In Fig. $14$ the resulting value of $\Delta \Omega$ is marked by a cross.
If $\theta_{Y}$ is increased further, upon decreasing 
$\epsilon_{B}/\epsilon_{A}$, the Wenzel state becomes 
unstable and the iteration process always converges to the Cassie state, 
irrespective of the initial conditions of the iterative determination
of the equilibrium densities.
                                                                              
This calculation was repeated for $\epsilon_{AB} = 0.84 \times \epsilon_{AA}$,
(keeping $c_{B} = 0.05$, $w = 7 \times \sigma$, and
$D = 4 \times \sigma$ fixed). For a value of $\epsilon_{A}$
corresponding to $\theta_{Y}(c_{B} = 0) \approx 112^{\circ}$
for the pure A liquid, for $\epsilon_{B}/\epsilon_{A} = 2.333$, 
and for a contact angle $\theta_{Y}(c_{B} = 0.05) \approx 90^{\circ}$
the Cassie state turns out to be unstable
and the only stable state is the Wenzel state, as reported in
Sec. \Rnum{4}.B.2. With keeping $\epsilon_{A} = 0.9834 \times k_{B}T$
fixed, upon decreasing $\epsilon_{B}/\epsilon_{A}$ to $2$, 
{\it i.e.,} $\theta_{Y}(c_{B} = 0.05) \approx 97^{\circ}$, 
the Cassie state becomes metastable whereas the Wenzel state 
remains stable; the resulting value of $\Delta \Omega$ is shown
by a filled circle in  Fig. $15$.
\begin{figure}[h]      
\vspace*{0.0cm}
\hspace*{0.0cm}\includegraphics[scale = 0.3]{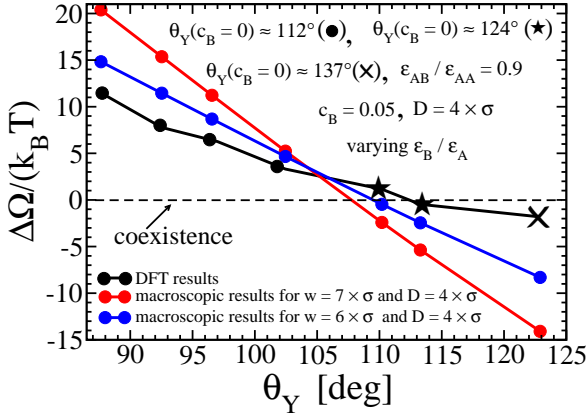}
{\vspace*{-0.0cm}
 \begin{spacing}{0.0}
\caption{Grand canonical free energy difference
 $\Delta \Omega = \Omega_{CB} - \Omega_{W}$ 
for the Cassie-Baxter and Wenzel states as  function 
of $\theta_{Y}(c_{B} = 0.05)$ for $\epsilon_{AB} = 0.9\times\epsilon_{AA}$ and 
$c_{B} = 0.05$, calculated both macroscopically (red 
and blue lines) and microscopically (black line).
We have chosen three values of $\epsilon_{A}$, resulting
in $\theta_{Y}(c_{B} = 0) \approx 112^{\circ}$
(shown by filled circles), $\theta_{Y}(c_{B} = 0) \approx 124^{\circ}$
(shown by stars), and $\theta_{Y}(c_{B} = 0) \approx 137^{\circ}$
(shown by a cross) for 
a pure A liquid. For each of these values of $\epsilon_{A}$,
$\epsilon_{B}/\epsilon_{A}$ is varied, represented in terms
of a change of $\theta_{Y}$. 
The pit geometry is characterized by
$w = 7\times \sigma$ and $D = 4\times \sigma$.}
 \end{spacing} }
\end{figure}

\begin{figure}[h]      
\vspace*{1.0cm}          
\hspace*{0.0cm}\includegraphics[scale = 0.3]{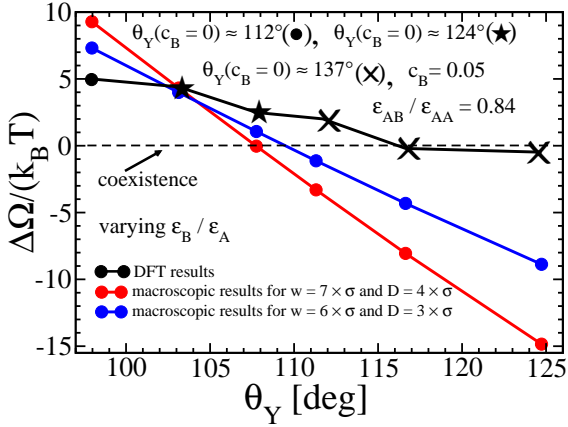}
{\vspace*{-0.0cm}
 \begin{spacing}{0.0}
\caption{Same as in Fig. $14$ but for $\epsilon_{AB} / \epsilon_{AA} = 0.84$.
Grand canonical free energy difference 
$\Delta \Omega = \Omega_{CB} - \Omega_{W}$ for the Cassie-Baxter and Wenzel
states as  function of $\theta_{Y}(c_{B} = 0.05)$
for $\epsilon_{AB} = 0.84\times\epsilon_{AA}$ and $c_{B} \approx 0.05$,
calculated both macroscopically (red and blue lines)
and microscopically (black line).
The blue line corresponds to the effective (reduced) pit dimensions.
We have chosen three values of $\epsilon_{A}$ corresponding to  
$\theta_{Y}(c_{B} = 0) \approx 112^{\circ}$ (filled circles), 
$\theta_{Y}(c_{B} = 0) \approx 124^{\circ}$ (stars),
and $\theta_{Y}(c_{B} = 0) \approx 137^{\circ}$
(crosses) for a pure A liquid. For each of these values of $\epsilon_{A}$, 
$\epsilon_{B}/\epsilon_{A}$ is varied, represented in terms 
of a change of $\theta_{Y}$. The pit geometry is characterized
by $w = 7\times \sigma$ and $D = 4\times \sigma$.
 }
 \end{spacing} }
\end{figure}
In the next step, we reduced $\epsilon_{A} = 0.8195 \times k_{B}T$ 
such that $\theta_{Y}(c_{B} = 0) \approx 124^{\circ}$
for the pure A liquid, and we considered 
two ratios, $\epsilon_{B}/\epsilon_{A} = 2.6$
and $2.4$, with corresponding contact angles 
$\theta_{Y}(c_{B} = 0.05) \approx 103^{\circ}$
and $\theta_{Y}(c_{B} = 0.05) \approx 107^{\circ}$
for the mixtures, respectively. The Cassie state 
remains metastable and the Wenzel state remains stable.
The corresponding values of $\Delta \Omega$ are indicated in Fig. $15$
by stars. The strength $\epsilon_{A}$ has been reduced further,
so that the corresponding contact angle $\theta_{Y}(c_{B} = 0)$, 
formed by the pure A liquid, increases to 
$\theta_{Y}(c_{B} = 0) \approx 137^{\circ}$.
Computations have been carried out for four ratios
$\epsilon_{B}/\epsilon_{A} = 3.25$, $3$, $2.5$,
and $2.375$. For $\epsilon_{B}/\epsilon_{A} = 3.25$
($\theta_{Y}(c_{B} = 0.05) \approx 111^{\circ}$ 
for the mixture) the Cassie state is still metastable and the Wenzel state
has still a lower grand canonical potential. For
$\epsilon_{B}/\epsilon_{A} = 3.0$ 
$(\theta_{Y}(c_{B} = 0.05)\approx 116^{\circ})$
both states coexist. The Wenzel state remains metastable
up to $\epsilon_{B}/\epsilon_{A} = 2.5$
$(\theta_{Y}(c_{B} = 0.05) \approx 124^{\circ})$. 
The value of $\Delta \Omega$  for this choice of $\epsilon_{A}$ 
(corresponding to $\theta_{Y}(c_{B} = 0.05) \approx 137^{\circ}$)
is indicated by a cross in Fig. $15$.
If  $\epsilon_{B}/\epsilon_{A}$ was decreased further to $2.375$, 
the Wenzel state became unstable.

\subsection{Influence of the pit dimensions}
Here, the influence of the pit dimensions on the Cassie - Wenzel transition
is discussed for the fluid-wall model studied in
Secs. \Rnum{4}.A.1 and \Rnum{4}.A.2.

\subsubsection{Varying the pit depth $D$}
\begin{figure}[h]      
\vspace*{0.10cm}
\hspace*{0.0cm}\includegraphics[scale = 0.3]{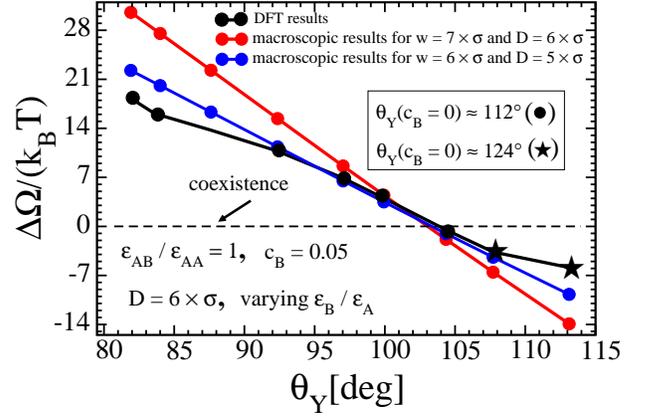}
{\vspace*{-0.0cm}
 \begin{spacing}{0.0}
\caption{$\Delta \Omega = \Omega_{CB} - \Omega_{W}$ as 
function of $\theta_{Y}$. The contact angle $\theta_{Y}(c_{B} = 0.05)$
is varied by changing $\epsilon_{A}$ and $\epsilon_{B}/\epsilon_{A}$. 
The black line shows DFT results, whereas the red and the blue 
line correspond to the macroscopic predictions.  We have 
chosen two values of $\epsilon_{A}$ resulting
in $\theta_{Y}(c_{B} = 0) \approx 112^{\circ}$ (filled circles) 
and $\theta_{Y}(c_{B} = 0) \approx 124^{\circ}$ (stars)
for a pure A liquid. For each of these two
values of $\epsilon_{A}$,  $\epsilon_{B}/\epsilon_{A}$ 
is changed in order to vary $\theta_{Y}(c_{B} = 0.05)$. 
The other parameters are fixed at $c_{B} = 0.05$, 
$\epsilon_{AA} = \epsilon_{AB} = \epsilon_{BB}$, 
$w = 7\times \sigma$, and $D = 6\times \sigma$.}
 \end{spacing} }
\end{figure}
In order to reveal the effect of the depth of the pit on the
Cassie - Wenzel transition, we have fixed the width
at $w = 7 \times \sigma_{A}$ whereas the depth $D$ of the pit is
changed from $4 \times \sigma$ to $6 \times \sigma$.
The other parameters are chosen as in Secs. \Rnum{4}.A.1 and
Sec. \Rnum{4}.A.2, {\it i.e.,} $\epsilon_{A}$
is taken such that the pure A liquid forms a contact angle of
$\theta_{Y}(c_{B} = 0) \approx 112^{\circ}$,
 with $\epsilon_{B}/\epsilon_{A}$
fixed at $2.833$ and $\epsilon_{AA} = \epsilon_{AB} = \epsilon_{BB}$.
Upon increasing $c_{B}$, as reported in
Sec. \Rnum{4}.A.1, for a depth of $4 \times \sigma$,
the Cassie state remains metastable up to $c_{B} \approx 0.128$
($\theta_{Y}(c_{B} \approx 0.128)\approx 83^{\circ}$
for the mixture).
The Cassie state becomes unstable for $c_{B} \approx 0.138$
($\theta_{Y}(c_{B} \approx 0.138) \approx 81^{\circ}$
for the mixture). However, for $D = 6 \times \sigma$, 
the Cassie state remains metastable up to the higher value of 
$c_{B} \approx 0.155$ which corresponds to a lower value 
$\theta_{Y}(c_{B} \approx 0.155) \approx 78^{\circ}$
of the mixture and becomes unstable for $c_{B} \approx 0.165$
($\theta_{Y}(c_{B} \approx 0.165) \approx 76^{\circ}$
for the mixture).

Equation (\ref{eq:18}), which is based on the macroscopic theory, predicts that the
critical contact angle $\theta^{Y}_{c}$, at which the CB and Wenzel
states in a pit of finite depth coexist and above which the CB
state turns into the stable state, shifts from 
$\theta^{Y}_c \approx 108^{\circ}$ for the pit depth
 $D = 4\times \sigma$ and a width of
$w = 7\times \sigma$, to $\theta^{Y}_{c} \approx 103^{\circ}$ if the
depth is increased to $D = 6\times \sigma$ while keeping the width fixed.
In order to test this prediction, the calculations of Sec. \Rnum{4}.A.2
for pits of a depth $D = 4\times \sigma$ have been repeated for deeper
pits with $D = 6\times \sigma$. As in Sec. \Rnum{4}.A.2, Young's contact
angle $\theta_{Y}$ is varied by changing the fluid-wall interactions
$\epsilon_{A}$ and $\epsilon_{B}$. Furthermore, the same width 
$w = 7\times \sigma$ of the pit, the same concentration $c_{B}$ of B
particles, and the same fluid-fluid interaction parameters
$\epsilon_{AA} = \epsilon_{AB} = \epsilon_{BB} = 0.9834\times k_{B}T$
have been chosen as in Sec. \Rnum{4}.A.2.
A set of calculations with two distinct values of $\epsilon_{A}$
have been carried out. For the first set, $\epsilon_{A}$ has been
chosen such that Young's contact angle $\theta_{Y}$ for a pure
A liquid is ca. $\theta_{Y}(c_{B} = 0) \approx  112^{\circ}$. 
The contact angle, for the given bulk concentration  $c_{B} =0.05$, 
is tuned by changing the ratio $\epsilon_{B}/\epsilon_{A}$. The chosen ratios are
$3$, $2.933$, $2.833$, $2.666$, $2.5$, $2.333$, $2.0$, and $1.666$
corresponding to values of $\theta_{Y}(c_{B} = 0.05)$ ranging from about 
$81^{\circ}$ to $107^{\circ}$. For $\epsilon_{B}/\epsilon_{A} > 2.933$, 
the Cassie state becomes unstable and the only stable state for the system
is the Wenzel state. For $\epsilon_{B}/\epsilon_{A} = 2.933$
($\theta_{Y}(c_{B} = 0.05) \approx 81^{\circ}$ for the mixture), 
the Cassie state becomes metastable and the Wenzel state is the stable one.
 The value of $\epsilon_{B}/\epsilon_{A}$ has been decreased
gradually, with keeping $\epsilon_{A}$ fixed, down to
$\epsilon_{B}/\epsilon_{A} = 1.666$.
For  $\epsilon_{B}/\epsilon_{A} = 2.0$
$(\theta_{Y}(c_{B} = 0.05) \approx 104^{\circ})$, 
the grand canonical potentials for the Cassie and the Wenzel states are
almost equal. For  $\epsilon_{B}/\epsilon_{A} = 1.666$
$(\theta_{Y}(c_{B} = 0.05) \approx 107^{\circ})$, the Wenzel state
becomes metastable and the Cassie state turns into the
stable state with the lower grand canonical potential.
Large contact angles are modeled by reducing $\epsilon_{A}$ to a
value corresponding to $\theta_{Y}(c_{B} = 0) \approx 127^{\circ}$
for a pure A fluid. Calculations have been carried out
for $\epsilon_{B}/\epsilon_{A} = 2.8$, $2.6$, and
$2.4$ corresponding to $\theta_{Y}(c_{B} = 0.05) \approx 110^{\circ}$,
$113^{\circ}$, and $117^{\circ}$ for the mixture. The Wenzel state remains
metastable up to $\epsilon_{B}/\epsilon_{A} = 2.8$
$(\theta_{Y}(c_{B} = 0.05) \approx 113^{\circ})$. 
If $\epsilon_{B}/\epsilon_{A}$ is decreased further, the Wenzel 
state becomes unstable. At $\theta_{Y}(c_{B} = 0.05) \approx 117^{\circ}$, 
the Cassie state is the stable state. The free energy differences $\Delta \Omega$
are shown in Fig. $16$ (DFT: black line). The results corresponding 
to the above first choice for $\epsilon_{A}$ are marked by filled 
circles, whereas the results corresponding to the second choice for 
$\epsilon_{A}$ are indicated by stars. The red and the blue lines correspond to macroscopic
predictions (Eq. (\ref{eq:17})) for the actual and the effective pit dimensions, 
respectively.
The effective quantities $w$ and $D$ are smaller than the actual ones
due to the presence of the depletion zone as well
as due to the slight intrusion of the liquid-vapor interface 
into the pit, even in the Cassie state. 
Within DFT, the shift of $\theta^{Y}_{c}$, 
as predicted by the macroscopic theory, is indeed observed (compare Figs. $8$ and $16$)
and, as predicted by the macroscopic theory, almost
quantitatively agrees with the DFT results. However,
the DFT values of $\Delta \Omega$ deviate considerably
from the macroscopic predictions.

\subsubsection{Influence of the width}

In order to study the influence of the width of the pits on
the stability and metastability of the Wenzel and the
Cassie states, we have fixed $c_{B} = 0.05$,
$D = 4\times \sigma$, $\epsilon_{AA} = \epsilon_{AB} = \epsilon_{BB}$
and decreased $w$ from $7\times \sigma$ to $5\times \sigma$. The value of
$\epsilon_{A}$  has been chosen such that the contact angle formed
by the pure A fluid is $\theta_{Y}(c_{B} = 0) \approx 112^{\circ}$.
Computations have been carried out for three distinct ratios
$\epsilon_{B}/\epsilon_{A} = 2.333, 2.0$, and $1.666$, corresponding to 
$\theta_{Y}(c_{B} = 0.05) \approx 99^{\circ}$,
$104^{\circ}$, and $107^{\circ}$ for the mixture. For the set of parameters 
leading to $\theta_{Y}(c_{B} = 0.05) \approx 99^{\circ}$,
the Wenzel state is the stable state and the Cassie state is
unstable. For $\theta_{Y}(c_{B} = 0.05) \approx 104^{\circ}$, 
the Cassie and the Wenzel states coexist. For 
$\theta_{Y}(c_{B} = 0.05) \approx 107^{\circ}$,
the Wenzel state becomes unstable and the Cassie state
turns into the stable state. We have repeated the same calculation 
for an even smaller width $w = 4\times \sigma$.
For this width, the Wenzel state is the stable state with
$\theta_{Y}(c_{B} = 0.05) \approx 99^{\circ}$. 
For the next tested value $\theta_{Y}(c_{B} = 0.05) \approx 104^{\circ}$, 
the Wenzel state becomes unstable and the Cassie state 
is the stable one.
No metastable states have been found for $w = 4\times \sigma$.
\section{ \bf Summary and Conclusions}
We have studied the Cassie - Wenzel transition of a symmetric 
binary liquid mixture at a nano-corrugated
surface which exhibits a periodic array of nanopits with a 
square cross section. The liquid is composed of two types
of particles labeled as A and B. The B
particles are taken to be attracted by the wall more strongly
than the A particles. The intrusion behavior of such
liquids has been studied as  function of
their composition, of the strength of the A $-$ B interaction
relative to the  A $-$ A and B $-$ B ones, of the relative
strengths of the A $-$ wall and the B $-$ wall interactions,
as well as of the pit dimensions.
Our study has been restricted to mixed liquids at liquid-vapor
coexistence. The structural properties of the binary liquid
mixtures in thermal equilibrium and, in case it
applies, in metastable equilibrium have been determined by
using density functional theory, which captures the microscopic
details of the system. The grand canonical free energy differences
$\Delta \Omega$ between the stable equilibria, {\it i.e.},
the Cassie or the Wenzel configuration and, as far as they occur,
the competing metastable equilibria, {\it i.e.,} the complementary Wenzel
or Cassie configuration, respectively, have been calculated as
functions of various system parameters. These results are
compared  with corresponding predictions from 
a macroscopic description in terms of surface tensions only.
We have found that intrusion of the liquid into the pits,
known as the Cassie $-$ Wenzel transition, cannot be predicted
reliably on the basis of a {\it single} parameter,
which is the contact angle, like the macroscopic theory
does. For instance, we have found that liquid intrudes a pit $-$ of
given geometry and fixed fluid-wall interactions $-$ at a variety of
contact angles if the liquid consists of symmetric binary fluids
with various ratios of the A $-$ B and A $-$ A interaction
strengths. Once, in the aforementioned case, 
the contact angle is reduced by
increasing the concentration of B particles, the liquid intrudes already
at a contact angle substantially higher than $ 90^{\circ}$ for a ratio of 
the 0.84 of the A $-$ B and A $-$ A interaction strengths, whereas, if this ratio
is chosen to be $1.0$, the contact angle has to be reduced
considerably below $ 90^{\circ}$ until liquid intrudes the pits.
With decreasing values of the ratio $\epsilon_{AB}/\epsilon_{AA}$,
the concentration of B particles increases at the liquid-vapor
interface (see Fig. $3$(d)). Since the wall $-$ B
attraction is stronger than the wall $-$ A attraction,
the wall might favor intrusion at contact angles above 
$90^{\circ}$ for decreasing values of 
$\epsilon_{AB}/\epsilon_{AA}$.
In such cases, the liquid-vapor interface moves into the pit having 
on its front side an excess of particles strongly
attracted by the walls. For the shallow pits studied here,
the contact angle, at which liquid intrudes, depends on the
depth of the pit. This contact angle is smaller
for deeper pits.

The free energy differences $\Delta \Omega$ between the
Cassie and the Wenzel configuration,  determined by using
density functional theory (DFT), deviate quantitatively
from the corresponding macroscopic predictions in terms
of surface tensions. An improved agreement between the
macroscopic predictions and the DFT results can be achieved,
if in the macroscopic expression for $\Delta \Omega$
a reduced effective width and a reduced effective depth of the
pits is introduced, which are introduced by taking depletion 
zones into account.
Nonetheless, substantial deviations persist. Typically,
but not always, the absolute values of the free energy differences
$\Delta \Omega$ computed by using DFT are smaller than those
obtained from the modified macroscopic expression.
This analysis is not restricted to nano-sized pits only, 
but also holds for binary colloidal suspensions 
intruding into micron-sized pits.

All computations presented here have been carried out at a 
fixed ratio $\epsilon_{AA}/k_{B}T = 0.9834$, corresponding
to temperatures well below the critical ones for all ratios  
$\epsilon_{AB}/\epsilon_{AA}$ studied. The influence
of temperature on the Cassie $-$ Wenzel 
transition is not studied explicitly, 
but only implicitly and to a certain extent,  
via the dependence of the transition on Young's 
contact angle, which is temperature dependent. 
More detailed investigations of this issue
are left to future studies. Furthermore, the 
present investigations have been limited 
to liquid-vapor coexistence. Pressures 
above the coexistence pressure tend to stabilize the Wenzel 
state. The analysis of pressure induced 
Cassie $-$ Wenzel transitions 
for binary liquid mixtures is also left to future studies.
\clearpage
\begingroup

\end{document}